\documentstyle[preprint,aps,epsf]{revtex}
\begin{document}
\draft

\title{
Perturbative study for domain-wall fermions in 4+1 dimensions}

\author{S. Aoki and H. Hirose}
\address{Institute of Physics, University of Tsukuba, Tsukuba
Ibaraki-305, Japan}

\date{\today}

\maketitle
\begin{abstract}
We investigate a U(1) chiral gauge model in 4+1 dimensions
formulated on the lattice via the domain-wall method.
We calculate an effective action for smooth background gauge fields
at a fermion one loop level. From this calculation
we discuss properties of the resulting 4 dimensional theory, 
such as gauge invariance of 2 point functions,  gauge anomalies
and an anomaly in the fermion number current.

\end{abstract}

\pacs{11.15Ha,11.30.Rd}
\narrowtext
\section{Introduction}
\label{sec:Int}

The standard model is very successful to explain many aspects
of electro-weak interactions.
However these successes come mainly from perturbative analysis,
and physics at the breaking scale, for example,  
the mass of the Higgs particle and the baryon number violation etc,
can not be predicted.
In order to predict them,
we need to study the standard model {\sl non-perturbatively}
-- especially using the technique of lattice gauge theories.

A main problem for studying the standard model on the lattice is the
difficulty to define lattice chiral gauge theories
due to the fermion doubling phenomenon\cite{NN,Kar}, which
can be easily seen in the fermion propagator on the lattice;
\begin{eqnarray}
S_F (p) = {\sum_\mu \gamma_\mu \sin(p_\mu) 
          \over \sum_\mu \sin^2 (p_\mu) }.
\end{eqnarray}
This propagator has poles at $p_\mu = (\pi,0,0,0)$, etc
as well as at $p_\mu = (0,0,0,0)$.
Therefore
a naively discretized lattice fermion field yields $2^d$ fermion modes, 
half of one chirality and half of the other, 
so that the theory is no more chiral
and therefore can not be used to construct the standard model on the lattice.
Several lattice approaches have been proposed to define chiral gauge theories,
but so far none of them have been proven to work successfully.

Recently Kaplan has proposed a new approach\cite{Kap} to this problem.
He suggested that 
it may be possible to simulate the behavior of massless chiral fermions 
in 2k dimensions 
by a lattice theory of massive fermions in 2k+1 dimensions 
if the fermion mass has a shape of a domain wall in the 2k+1-th 
dimension. 
He showed for the weak gauge coupling limit
that the massless chiral states arise as zero-modes bound to
the 2k-dimensional domain wall 
while all doublers can be given large gauge invariant masses. 
If the chiral fermion content that appears 
on the domain wall 
is anomalous the 2k-dimensional 
gauge current flows off the wall into
the extra dimension so that the theory 
can not be 2k-dimensional.
Therefore he argued that
this approach possibly simulates the 2k-dimensional chiral 
fermions only for anomaly-free cases.  

His idea, called a domain-wall fermion method, 
was tested for smooth external gauge fields.
It has been shown both numerically\cite{Jan} and analytically\cite{Our} 
that in the case of the chiral Schwinger model the anomaly 
in the gauge current is cancelled on the wall 
among three fermions of charge 3, 4, and 5. 
The Chern-Simons current was also evaluated in Ref.\ \cite{GJK,Our}:
It is shown that the 2k+1-th component of the current is non-zero
in the positive mass region and zero in the negative mass region,
so that the derivative of the current cancels 
the 2k-dimensional gauge anomaly on the wall, as was argued in Ref.\ 
\cite{Kap,Kap2}
\footnote{Recently the anomaly is also calculated
in the continuum version of the domain-wall fermion\cite{KiKu,Candra,RS}.}.

Results above provide positive indications that the domain-wall fermion
method may work as a lattice regularization for chiral gauge theories.
There exists, however, two remaining problems to be considered in 
this approach.
One of the problems is the fate of the chiral zero mode on the domain wall:
Since the original 2k+1-dimensional  model is vector-like,
there always exists an anti-chiral mode, localized on an anti-domain wall
formed by periodicity of the extra dimension.
If the chiral mode and the anti-chiral mode are paired into
a Dirac mode, this approach fails to simulate chiral gauge theories.
Without dynamical gauge fields, the overlap between the chiral mode and 
the anti-chiral mode is suppressed as $O (e^{-L})$ where $L$ is the size of the
extra dimension. If gauge fields become dynamical, the overlap
depends on the gauge coupling.
It was found\cite{Alt,Todai} that the chiral mode disappears 
and the model becomes vector-like
in the strong gauge coupling limit of the extra dimension. 
Recently this problem has been investigated at the intermediate coupling
region via the numerical simulation for  a 2+1 dimensional U(1) model, but 
no definite conclusion on the existence of the chiral zero mode
can be obtained in the symmetric phase\cite{AN}.

The other problem is related to a structure of an effective action
for smooth back-ground gauge fields at the fermion 1-loop level:
The perturbative evaluation for the 2+1 dimensional model found\cite{Our} that,
if gauge fields depend on coordinates of the extra dimension,
the effective action contains the longitudinal component as well as parity-odd
terms, and that this longitudinal component, which breaks gauge invariance,
remains nonzero even for anomaly-free cases. The gauge non-invariant 
parity-even term
seems absent\cite{AL,Sham} in two modifications of the Kaplan's
original domain wall fermion, 
wave-guide model\cite{GJPV,GoSha} and overlap formula\cite{NarNeu,Nar}, 
whose gauge fields do not
depend on coordinates of the extra dimension. These two modifications, however,
suffers from the first problem: the chiral zero mode on the domain wall 
seems to disappear in the presence of the dynamical gauge 
fields\cite{GJPV,GoSha,GoSha2}.

In this paper, following the previous calculation in 2+1 dimensions\cite{Our},
we have carried out a detailed perturbative calculation of the original
domain wall fermion formulation in 4+1 dimensions 
for smooth background gauge fields,
in order to investigate the structure of the effective action in higher
dimensions. 
In sect.\ \ref{sec:For},
we briefly summarize the lattice perturbation theory for
the domain wall method with the periodic boundary condition\cite{Our}.
In sect.\ \ref{sec:41},
we evaluate the 2-point function and anomaly in 4+1 dimensions
at a fermion one-loop order. 
We find that the effective action for the 4+1 dimensional theory has 
the similar structure to that for the 2+1 dimensional theory:
there appear
not only parity-odd terms such as the gauge anomaly and the Chern-Siomns term
but also parity even terms such as the mass term and the Lorentz non-covariant
term. Therefore the gauge non-invariant terms remin non-zero 
for anomaly free cases also in 4 dimensions. 
In sect.\ \ref{sec:PI},
we comment on an anomaly of the fermion number current
in 4+1 dimensions.
Finally we give our conclusions in sect.\ \ref{sec:Con}.

\section{Formulation}
\label{sec:For}

In this section,
we briefly summarize a formulation of the domain wall method and
a set-up of lattice perturbation theories. In particular, we explicitly
give a fermion propagator, vertex functions, and the Ward-Takahashi identity.
\subsection{Lattice Action}

We consider a vector gauge theory in D=2k+1 dimensions with
a domain wall mass term.  For later convenience we use the
notation of Ref.\ \cite{NarNeu}, 
where the fermionic action is written in terms of
a 2k dimensional theory with infinitely many flavors.
Our action is denoted as

\begin{equation}
S= S_G + S_F .
\end{equation}
The action for gauge field $S_G$ is given by
\begin{eqnarray}
S_G & =& 
\beta \sum_{n,\mu>\nu}\sum_s {\rm Re}\{ {\rm Tr} [U_{\mu\nu}(n,s)]\}
	\nonumber \\
& +& \beta_D \sum_{n,\mu}\sum_s {\rm Re} \{ {\rm Tr} [U_{\mu D}(n,s)]\}
\end{eqnarray}
where $\mu$, $\nu$ run from 1 to $2k$, 
$n$ is a point on a 2k-dimensional lattice and $s$  a coordinate 
in the extra dimension,
$\beta$ is the inverse gauge coupling for plaquettes $U_{\mu\nu}$
and $\beta_D$  that for plaquettes $U_{\mu D}$.
The fermionic part of the action $S_F$ is given by

\begin{eqnarray}
S_F & = & {1\over 2}\sum_{n,\mu}\sum_s  \bar\psi_s(n)\gamma_\mu
[U_{s,\mu}(n)\psi_s(n+\mu ) \nonumber  \\
&& \qquad \qquad \qquad \qquad \qquad
- U^\dagger_{s,\mu}(n-\mu)\psi_s(n-\mu ) ] 	\nonumber \\
&+&  \sum_n \sum_{s,t} \bar\psi_s(n) [ M_0 P_R + M_0^\dagger P_L]_{st}
\psi_t(n)  
\label{eqn:actionf} \\
 & + & {1\over 2}\sum_{n,\mu}\sum_s  \bar\psi_s(n)
[U_{s,\mu}(n)\psi_s(n+\mu )\nonumber  \\
 &&  \qquad \qquad + U^\dagger_{s,\mu}(n-\mu)\psi_s(n-\mu ) 
    -2\psi_s(n) ] \nonumber 
\end{eqnarray}
where $s$, $t$ are considered as flavor indices, 
$P_{R/L} = (1 \pm \gamma_{2k+1})/2$,

\begin{eqnarray}
(M_0)_{st} & = & 
U_{s,D}(n)\delta_{s+1,t}- a(s)\delta_{st}   \nonumber	\\ 
(M_0^\dagger)_{st} & = & 
U^\dagger_{s-1,D}(n)\delta_{s-1,t}- a(s)\delta_{st} ,
\end{eqnarray}
and $U_{s,\mu}(n)$, $U_{s,D}(n)$ are link variables for gauge fields.
We consider the above model with a periodic boundary
in the extra dimension,
so that $s$, $t$ run from $-L$ to $L-1$, and we take
\begin{eqnarray}
a(s) & =& 
1 - m_0[ {\rm sign}(s+{1\over 2}) \cdot {\rm sign}(L-s-{1\over 2})]
\nonumber \\
     & =& \left\{ 
\begin{array}{ll}
1-m_0, &  -{1\over 2} < s < L-{1\over 2} \\
1+m_0, & -L-{1\over 2} < s < -{1\over 2}
\end{array}
\right.
\end{eqnarray}
for $-L \leq s < L $.  It is easy to see\cite{NarNeu} that  $S_F$ above is 
identical to the domain-wall fermion action in D=2k+1 dimensions\cite{Kap}
with the Wilson parameter $r=1$. In fact the second term 
in eq.\ (\ref{eqn:actionf}) can be rewritten as
\begin{eqnarray}
& & {1\over 2} \bar\psi_s\gamma_D [U_{s,D}\psi_{s+1}-U_{s-1,D}\psi_{s-1}] 
	\nonumber \\
&+& {1\over 2}\bar\psi_s[U_{s,D}\psi_{s+1}+U_{s-1,D}\psi_{s-1}
-2\psi_s] + M(s) \bar\psi_s \psi_s  
\end{eqnarray}
with $M(s)=m_0 [ {\rm sign}(s+1/2) \cdot {\rm sign}(L-s-1/2)]$.
It is easy to see that one chiral zero mode appears on the domain wall
around $s=0$ if and only if $0 < m_0 < 2$.

\subsection{Fermion propagator and Feynman rules}

The fermion propagator in 2k-dimensional momentum space and
in real D-th space has been obtained in Ref.\ \cite{NarNeu,Our} for
large $L$:
\begin{eqnarray}
S_F(p)_{st} = - [ [(i\sum_\mu \gamma_\mu \bar p_\mu + M)G_L(p)]_{st}P_L 
\nonumber \\
+ [(i\sum_\mu \gamma_\mu \bar p_\mu + M^\dagger)G_R(p)]_{st}P_R ] 
\label{eqn:fprop} 
\end{eqnarray}
where
\widetext
\begin{equation}
G_L(p)_{st}   = 
\left\{ \begin{array}{l}
(s,t \ge 0) \\
\qquad B e^{-\alpha_+|s-t|} 
+ (A_L-B) e^{-\alpha_+(s+t)} + (A_R-B) e^{-\alpha_+(2L-s-t)}, \\
(s\ge 0,\ t\le 0) \\
\qquad A_Le^{-\alpha_+s+\alpha_-t} +A_Re^{-\alpha_+(L-s)-\alpha_-(L+t)},\\
(s\le 0,\ t\ge 0) \\
\qquad A_Le^{\alpha_-s-\alpha_+t} + A_Re^{-\alpha_-(L+s)-\alpha_+(L-t)}, \\
(s,t \le 0) \\
\qquad C e^{-\alpha_-|s-t|} 
+ (A_L-C) e^{\alpha_-(s+t)} + (A_R-C) e^{-\alpha_-(2L+s+t)},
\end{array}  \right. 
\label{eqn:gl}
\end{equation}
\begin{equation}
G_R(p)_{st}  = 
\left\{ \begin{array}{l}
 (s,t \ge -1) \\
\qquad B e^{-\alpha_+|s-t|}  + (A_R-B) e^{-\alpha_+(s+t+2)}  \\
\qquad \qquad \qquad \qquad  \qquad \qquad \qquad 
+(A_L-B) e^{-\alpha_+(2L-s-t-2)},  \\
 (s\ge -1,\  t\le -1) \\
\qquad A_Re^{-\alpha_+(s+1)+\alpha_-(t+1)} +A_Le^{-\alpha_+(L-s-1)-\alpha_-(L+t+1)}, \\
 (s\le -1,\  t\ge -1) \\
\qquad A_Re^{\alpha_-(s+1)-\alpha_+(t+1)} +A_Le^{-\alpha_-(L+s+1)-\alpha_+(L-t-1)},\\
 (s,t \le -1) \\
\qquad C e^{-\alpha_-|s-t|} +  (A_R-C) e^{\alpha_-(s+t+2)} \\ 
\qquad \qquad \qquad \qquad  \qquad \qquad \qquad 
+(A_L-C) e^{-\alpha_-(2L+s+t+2)},  \\
\end{array}  \right. 
\label{eqn:gr}
\end{equation}
\narrowtext
with
\begin{eqnarray}
a_{\pm} &=& 1 -{\nabla (p)\over 2}\mp m_0 \\
\alpha_{\pm} &=&  
{\rm arccosh} [{1\over 2}(a_{\pm}+{1+\bar p^2\over a_{\pm}})] \geq 0, \\
A_L  &=& {1\over a_+  e^{\alpha_+} - a_-e^{-\alpha_-}},
\label{eqn:ALR} \\
A_R &=&{1\over a_-  e^{\alpha_-}- a_+e^{-\alpha_+}} , \\
B &=& {1\over 2 a_+ {\rm sinh} \alpha_+}, \\
C &=& {1\over 2 a_- {\rm sinh} \alpha_-}. 
\end{eqnarray}
 
From the form of $A_L$, $A_R$, $B$ and $C$,
it is easy to see 
that singularities occur only in $A_L$ at $p=0$;
\begin{equation}
A_L \longrightarrow {m_0(4-m_0^2) \over 4 p^2 a^2 },
\qquad p\rightarrow 0  .
\label{eqn:singK}
\end{equation}
Therefore the $G_L$ part of the propagator describes one massless right-handed
fermion around $s,t = 0$, which corresponds to the zero mode on the domain 
wall.
It is also noted that the $G_R$ part describes one 
massless left-handed fermion around $|s|,|t| = L$,
which corresponds to the anti-zero mode, 
due to the periodic boundary condition in the extra dimension.

\

Now
we write down the lattice Feynman rules 
relevant for a fermion one-loop calculation, 
which will be performed in the next section.
We first choose the axial gauge fixing\footnote{
To choose the axial gauge fixing in the periodic boundary condition,
gauge field configurations should satisfy a constraint 
that the Polyakov loop in the extra dimension is equal to unity.
To achieve this constraint, we should put a delta function of the
constraint, so that the other gauge coupling $g_s \propto 1/\sqrt{\beta_s}$
is not necessarily small and can be made arbitrary large,
or we should also take the weak coupling limit of $g_s$.}
: $U_{s,D}=1$.
Although the full gauge symmetries in D dimensions are lost,
the theory is still invariant under 
$s$-independent gauge transformations \cite{Dist}.  
Therefore the gauge current $J_\mu (x) = \sum_s j_\mu (x,s)$ is conserved.
We consider the limit of small 2k-dimensional gauge coupling, and take
\begin{equation}
U_{s,\mu}(n) = \exp [i a g A_\mu (s, n+\mu/2)]
\end{equation}
where $a$ is the lattice spacing, and $g \propto 1/\sqrt\beta$  
is the gauge coupling constant whose mass dimension is $2-D/2$
(mass dimension of the gauge fields $A_\mu$ is $D/2-1$).
We consider Feynman rules in momentum space 
for the physical 2k dimensions 
but in real space for the extra dimension. There are three relevant points
for later calculations.

The fermion propagator $ S_F(p)_{st}$ has been given 
in eq.(\ref{eqn:fprop}) 
with eqs.(\ref{eqn:gl}, \ref{eqn:gr}). 

The fermion vertex coupled to  a single gauge field is given by
\begin{eqnarray}
ag  \partial_\mu [ S_F^{-1}(k)]_{st} 
 = i \cos (a k_\mu)\gamma_\mu \delta_{s,t} + \sin(a k_\mu)\delta_{s,t}
\label{eqn:v1}
\end{eqnarray}
where $\displaystyle k ={q+p\over 2}$ and
the fermion vertex with two gauge fields is
\begin{eqnarray}
-a^2\frac{g^2}{2} \partial_\mu^2 [ S_F^{-1}({q+p\over 2})]_{ss}.
\label{eqn:v2}
\end{eqnarray}
From the periodicity of eq.(\ref{eqn:v1}),
the fermion vertex with $2n+1$ gauge fields is proportional to 
$ \partial_\mu [ S_F^{-1}({q+p\over 2})]_{st} $
and 
the fermion vertex with $2n$ gauge fields is proportional to 
$ \partial_\mu ^2 [ S_F^{-1}({q+p\over 2})]_{st} $.

The fermion propagator satisfies the Ward-Takahashi identity on the lattice.
The identity is given by
\begin{eqnarray}
\sum_{\mu =1} ^{2k} 
2 \sin(p_\mu /2) \partial_\mu [ S_F^{-1}({q+p/2})]_{s,t}
&=&[S_F^{-1}({q+p})]_{s,t} \nonumber \\
&-&[S_F^{-1}({q})]_{s,t}.
\label{eqn:WI}
\end{eqnarray}

\section{U(1) chiral gauge theory in 4+1 dimensions} 
\label{sec:41}

In this section we investigate an U(1) chiral gauge theory in 4+1 dimensions.
In 4 dimensions,
it is shown with the power counting that the n-point functions 
which has divergent diagrams are 2-,3- and 4-point functions.  
In the following two subsections we calculate in detail
the 2-point function of the gauge field $\Pi_{\mu \nu} (p)$ 
and the parity odd part of the 3-point function ( the gauge anomaly ).

\subsection{Calculation of 2-point function}

First of all, the effective action with two external gauge fields 
is denoted by
\begin{eqnarray}
S_{eff}^{(2)} & \equiv & -{g^2\over 2}\sum_{p,s,t}
A_\mu (s,p) A_\nu (t,-p) \Pi^{\mu\nu}(p)_{st}
	\nonumber \\
&=& -{g^2\over 2}\sum_{p,s,t} A_\mu (s,p) A_\nu (t,-p)
 [\Pi_a^{(2)} + \Pi_b^{(2)}]^{\mu\nu} _{st},
\end{eqnarray}
where 
\widetext
\begin{eqnarray}
[\Pi_a^{(2)}]^{\mu\nu} _{st}  & = & 
\int_{-\pi/a}^{\pi/a} {d^4 q\over (2\pi)^4}
{\rm tr} \left\{[\partial_\mu S_F^{-1}(q+{p\over 2})\cdot S_F(q+p)]_{st}
[\partial_\nu S_F^{-1}(q+{p\over 2})\cdot S_F(q)]_{ts}\right\}\times a^2, 
\end{eqnarray}
\begin{eqnarray}
[\Pi_b^{(2)}]^{\mu\nu} _{st}  & = & 
-\delta_{st} \delta_{\mu\nu} \int_{-\pi/a}^{\pi/a} {d^4 q\over (2\pi)^4}
{\rm tr} [\partial_\mu^2 S_F^{-1}(q)\cdot S_F(q)]_{ss} \times a^2 . 
\end{eqnarray}

This integral has the similar form as in the 2+1 dimensions case,
but has the divergence of order $a^{-2}$.
To separate the would-be divergent part from the finite part
we rewrite this integral as follows\cite{KN}:
\begin{eqnarray}
\Pi_{\mu \nu} (p) &=& 
 [ \Pi_{\mu \nu} (p) -  \Pi_{\mu \nu} (0) 
-{p_\rho p_\sigma \over 2} \partial_{\rho,\sigma} ^2 \Pi_{\mu \nu} (0)]
+[ \Pi_{\mu \nu} (0) 
   +{p_\rho p_\sigma \over 2} \partial_{\rho,\sigma} ^2 \Pi_{\mu \nu} (0)].
\end{eqnarray}
\narrowtext
The first derivative term disappears due to the symmetry of the integral. 
Note that 
we adopt the dimensional regularization with $4+2\epsilon$ dimensions to
avoid infra-red singularities for zero external momentum.
With this infra-red regularization the first term has already been finite,
so that it can be evaluated as 
the value of the naive continuum limit ($a \rightarrow 0$).
Thus we obtain,
\widetext
\begin{eqnarray}
\Pi_{\mu \nu} (p) &=&
 [ \Pi_{\mu \nu} ^{cont,}(p)   -\Pi_{\mu \nu} ^{cont.} (0) 
  -{p_\rho p_\sigma \over 2} 
       \partial_{\rho,\sigma} ^2 \Pi_{\mu \nu}  ^{cont.} (0)] 
+[ \Pi_{\mu \nu} (0) 
   +{p_\rho p_\sigma \over 2} \partial_{\rho,\sigma} ^2 \Pi_{\mu \nu} (0)]
\nonumber \\
&=& [ \Pi_{\mu \nu} ^{cont.}(p)]
   +[ \Pi_{\mu \nu} ^{lattice}(0) 
     +{p_\rho p_\sigma \over 2} 
          \partial_{\rho,\sigma} ^2 
               \Pi_{\mu \nu} ^{lattice}(0)],
\nonumber
\end{eqnarray}
\narrowtext
where $cont.$ stands for continuum.
In the last equality,
we use the fact that
the integral with zero external momenta is zero in the continuum dimensional 
regularization.
It is noted that
the ``cont. '' term is integrated 
with the dimensional regularization 
but the second term must be integrated on the lattice in $4+2\epsilon$
dimensions.
In other words, the first term is 
the contribution of the continuum theory and 
the second term is that of the lattice theory, and
therefore the latter is named as ``lattice''.
It is also noted that we use $\gamma_5$ which anti-commutes with all 
$\gamma_\mu$ in $4+2\epsilon$ dimensions for the calculation of 2-point
function. Since the final result is independent of the infra-red regulator,
so that it does not depend on the choice of $\gamma_5$.

\

\paragraph{Evaluation for the Continuum part}\

The continuum part leads to the usual transversal form, 
multiplied by the function $F_{L/R} (s,t)$ 
which characterizes the domain-wall fermion.
The result is
\widetext
\begin{eqnarray}
\Pi^{cont.} _{\mu\nu,st} (p) &=&
- g^2  \sum_\chi F^2 _\chi (s,t)  
\int_{-\infty} ^{\infty} {d^n q\over (2\pi)^n} 
{\rm tr} \left\{
i \gamma_\mu P_\chi {-i ({q\!\!\!\!/\,\,}+{{p} \!\!\!\!/\,\,}) \over (q+p)^2}
i \gamma_\nu P_\chi {-i {q \!\!\!\!/\,\,}\over q^2 } \right\} 
\nonumber \\
&=& {- g^2 \over (4\pi)^2} \sum_\chi F^2 _\chi 
\left[ {1 \over \epsilon } + {5 \over 3} -\gamma_E 
     + \log({4 \pi \bar \mu^2 \over p^2})\right]  {2 \over 3} T_{\mu\nu},
\end{eqnarray}
\narrowtext
where $\epsilon = \displaystyle {n-4 \over 2}$ and 
      $\bar \mu$ is the renormalized point 
and $T_{\mu\nu}=T_{\mu\nu}(p)$ is the transverse function of $p$:
$T_{\mu\nu}(p)= p^2 \delta_{\mu\nu} - p_\mu p_\nu$.

\

\paragraph{Evaluation for the Lattice part}\

The Lattice part $\Pi^{lattice} _{\mu\nu} (p)$ can be divided 
into two quantities,
a mass term of the gauge field 
and a second derivative term of the $\Pi_{\mu\nu}$,
which contains the transverse part. 

We consider the mass term 
$\Pi ^{lattice} _{\mu\nu ,st}(0) = \Pi_M(s,t) \delta_{\mu\nu}$, 
which has the following form :
\widetext
\begin{eqnarray}
\Pi_M(s,t)&=& - g^2 \int_{-\pi}^{\pi} {d^n q\over (2\pi)^n}{\rm tr} 
\left\{[\partial_\mu S_F^{-1}(q)\cdot S_F(q)]_{st}
[\partial_\mu S_F^{-1}(q)\cdot S_F(q)]_{ts}\right\}\times a^{-2}
\nonumber \\
&+& g^2 \delta_{st} 
\int_{-\pi}^{\pi}{d^n q\over (2\pi)^n}{\rm tr} 
[\partial_\mu^2 S_F^{-1}(q)\cdot S_F(q)]_{ss} \times a^{-2}.
\end{eqnarray}
Here  no sum over $\mu$ is taken.
By the rescaling $q \rightarrow q a $ and the fact
that the integral is infra-red finite in 4 dimensions,
we obtain
\begin{eqnarray}
\Pi_M(s,t) &=& 
g^2 a^{-2}  \int_{-\pi}^{\pi} {d^4 q\over (2\pi)^4}{\rm tr} 
\left\{\partial_\mu [L_\mu (q)]_{s,t}\right\} 
\\ &-& 
g^2 a^{-2} \int_{-\pi}^{\pi}{d^4 q\over (2\pi)^4}{\rm tr} 
\left\{ [L_\mu (q) \star L_\mu (q)]-[L_\mu (q) \cdot L_\mu (q)]
\right\} _{s,t}, 
\end{eqnarray}
where
\begin{eqnarray}
[L_\mu (q) ]_{s,t} &\equiv &[\partial_\mu S_F^{-1}(q)\cdot S_F(q)]_{st}, 
\end{eqnarray}
and
\begin{eqnarray}
[A \star    B ]_{s,t} = A_{s,t}  B_{t,s}, \qquad
[A \cdot B ]_{s,t} =  \delta_{s,t} \sum_u A_{s,u} B_{u,s}.
\end{eqnarray}
Since the first term is equal to zero because of the Stokes' theorem,
the mass term of the gauge field finally becomes
\begin{eqnarray}
\Pi_M(s,t) = 
&& - g^2 a^{-2} \int_{-\pi}^{\pi}{d^4 q\over (2\pi)^4}
{\rm tr} \left\{
[L_\mu (q) \star L_\mu (q)]-[L_\mu (q) \cdot L_\mu (q)]
\right\}_{s,t}. \nonumber 
\end{eqnarray}

\narrowtext
It is easy to check that
this mass term satisfies the following identity:
\begin{eqnarray}
\sum_{t=-L} ^L \Pi_M (s,t) = \sum_{s=-L} ^L \Pi_M (s,t)=0,
\label{eqn:2pmass}
\end{eqnarray} 
which comes from the fact that the theory with the axial gauge fixing
is still invariant under $s$ independent gauge transformations.
Thus, for the $s$ independent background gauge field $A_\mu (s,p)=
A_\mu (p)$, no mass term is generated by the fermion 1 loop integral.

\begin{figure}
\centerline{\epsfxsize=12cm \epsfbox{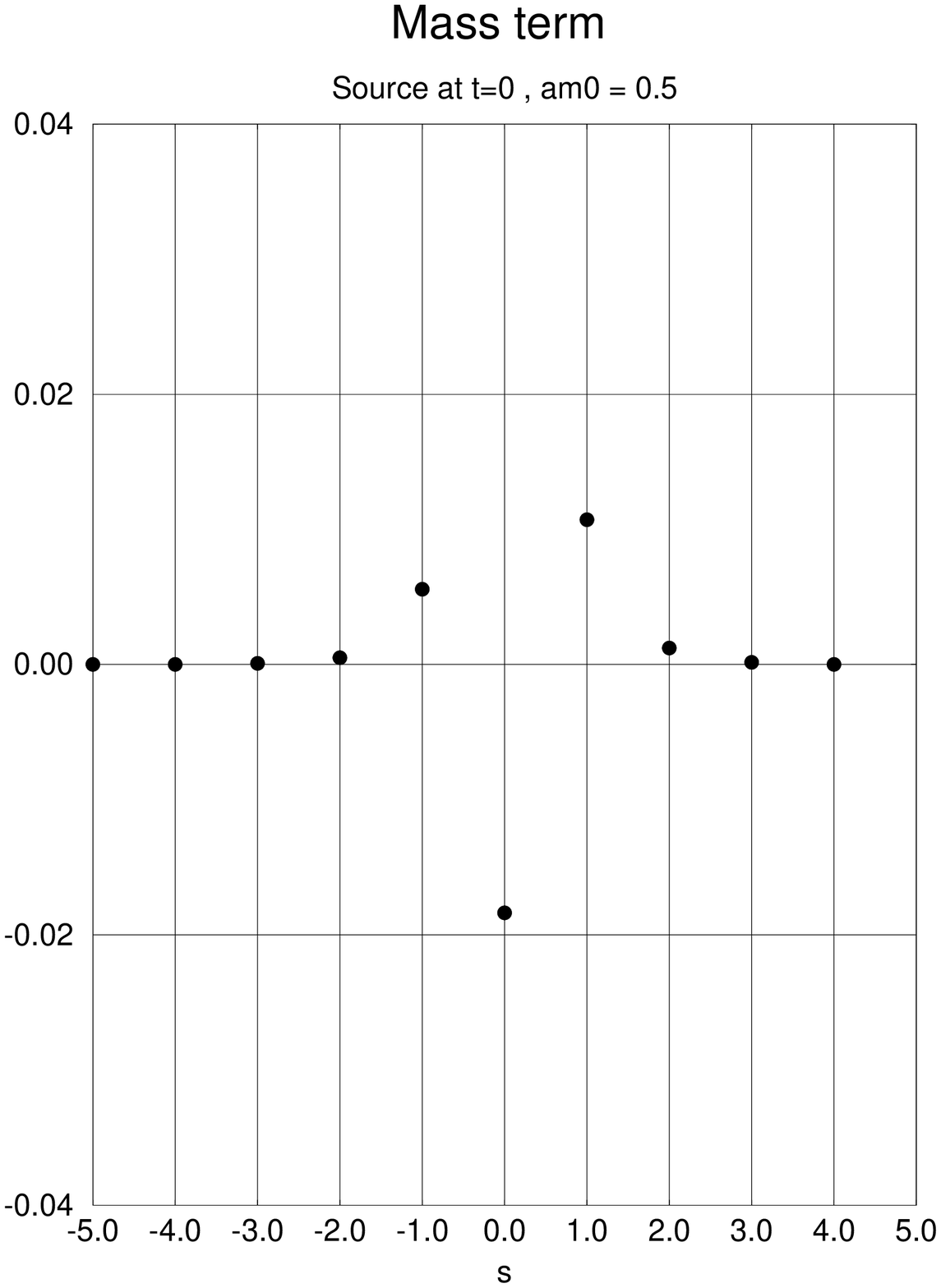}}
\caption{The coefficient of the 2 point function
$\Pi_M(s,t)$ as a function of $s$ 
with $t=0$ fixed at $m_0=0.5$ and $L=20$.}
\label{mass4st}
\end{figure}

\begin{figure}
\centerline{\epsfxsize=12cm \epsfbox{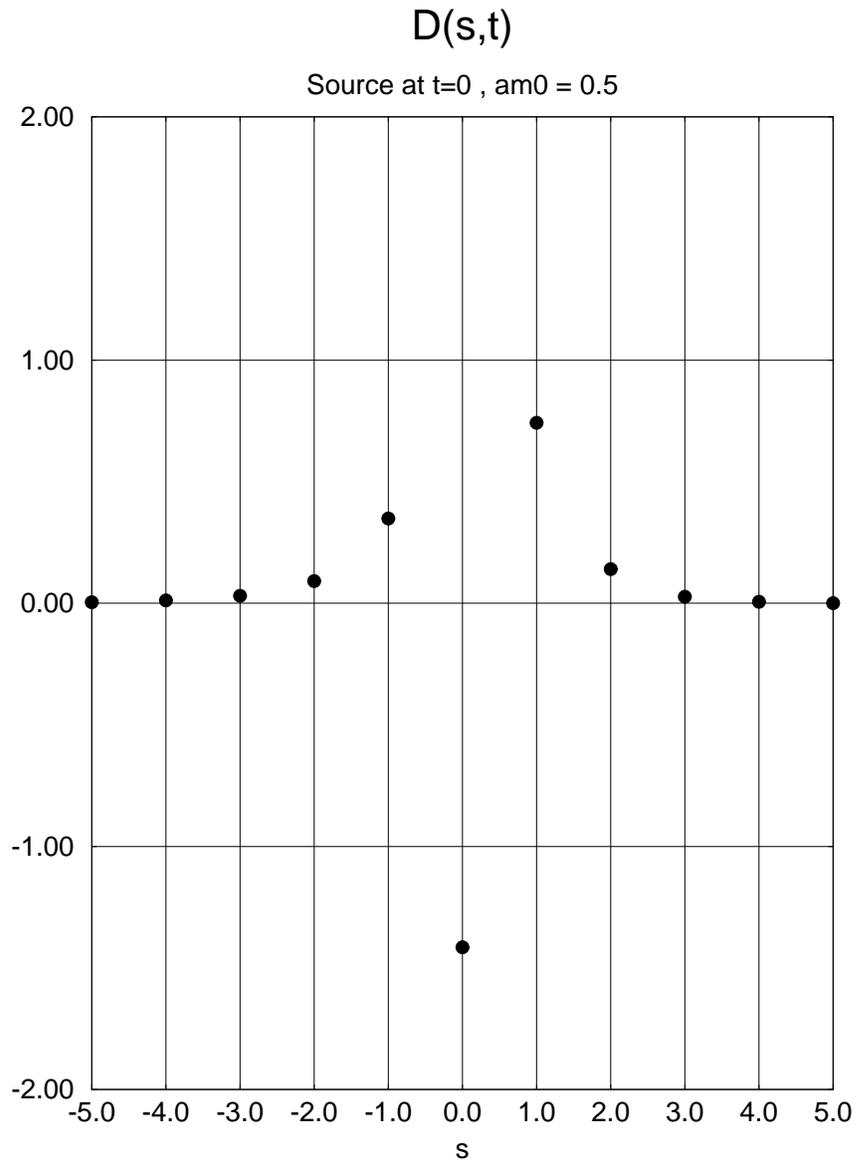}}
\caption{The coefficient of the even term $D(s,t)$
in 2+1 dimensions as a function of $s$ with $t=0$ fixed
at $m_0=0.5$ and $L=5$.}
\label{mass2st}
\end{figure}

Since it is difficult to calculate the mass term analytically for 
general $s$ and $t$, we evaluate it numerically.
In Fig.\ref{mass4st}, the behavior of the mass term $\Pi_M(s,t)$
with $t=0$ fixed is plotted as a function of $s$ at $m_0 = 0.5$
and $L=5$.
The $\Pi_M(s,0)$ has the largest (negative) values at $s=0$, the place where
a chiral zero mode lives. This means that a loop of the chiral zero mode 
mainly contributes to $\Pi_M(s,t)$. Furthermore we have checked
that $\Pi_M(s,s)$ is  small at $s\not=0$ or $\not=L$, where only massive modes 
exist. 
The behaviour of $\Pi_M(s,0)$ is similar to the shape of the mass term in 2+1 
dimensions\cite{Our}, which is given in Fig.\ref{mass2st}
where $D(s,0)$ is plotted as a function of $s$ at $m_0 = 0.5$
and $L=5$. 
Here $D(s,t) = \sum_X F_X(s,t)^2 - 2 K(s,t)$ and $K(s,t)$ is given in
ref.\cite{Our}.

Next let us consider the calculation of  the second derivative term.
Since the transversality is hidden behind this term,
we can parametrize $\displaystyle {p_\rho p_\sigma \over 2} 
\partial_{\rho,\sigma} ^2 \Pi_{\mu \nu} ^{lattice}(0)$ as follow. 
\widetext
\begin{eqnarray}
{p_\rho p_\sigma \over 2} 
\partial_{\rho,\sigma} ^2 \Pi_{\mu \nu} ^{lattice}(0) 
&=& \Pi^{(a)} T_{\mu\nu}(p) + \Pi^{(b)} \delta_{\mu\nu} p^2
+\Pi^{(c)} \delta_{\mu\nu} p_\nu ^2.
\label{eqn:para}
\end{eqnarray}
The second and the last term in the above equation 
break the transversality.
Especially, the last term breaks the Lorenz invariance as in the case of
Wilson-Yukawa formulation for lattice chiral gauge theories\cite{Aoki}.
Although the details of calculation given in Appendix B is important,
we give only the results of $\Pi ^{(a,b,c)}$ below, where $\alpha$ and 
$\beta $ are not equal and no sum over them is taken.
\begin{eqnarray}
\Pi^{(a)} =& -& \Pi^{(b)}+{g^2 \over 2}
\int_{-\pi}^{\pi}{d^n q\over (2\pi)^n} 
{\rm tr} \left\{
  [(L_\alpha (q) \cdot L_\beta   (q)) \star
   (L_\beta  (q) \cdot L_\alpha  (q))]
\right.\nonumber \\
& & \left. \qquad \qquad 
- [(L^{con} _\alpha (q) \cdot L^{con} _\beta   (q)) \star
   (L^{con} _\beta  (q) \cdot L^{con} _\alpha  (q))]
\right\}_{s,t} \nonumber \\ 
&-&{g^2 \over (4 \pi)^2 } \sum_{\chi=L,R} F^2 _\chi (s,t)
\left(
{1 \over 3}-{2 \over \sqrt{3}\pi } + {2 \times 0.46349 \over 3 \pi^2}
\right) \nonumber \\
&
+& {g^2 \over (4 \pi)^2 } \sum_{\chi=L,R} F^2 _\chi (s,t)
{2 \over 3}
 \left( {1 \over \epsilon} + log (a^2 \bar \mu^2)\right),
\label{eqn:2point-a}
\end{eqnarray}
where 
$L^{con} _\mu(q)$ is a value of $L_\mu(q)$ in the naive continuum limit.
Suppressing the extra dimension indices $s,t$, $\Pi^{(b)}$ becomes 
\begin{eqnarray}
\Pi^{(b)} &=& 
{-g^2 \over 4}
\int_{-\pi}^{\pi}{d^n q\over (2\pi)^n}{\rm tr} 
\left\{
L_\alpha (q) \cdot L_\alpha (q) \cdot
[L_\beta (q)\star 1 - 1 \star L_\beta (q)] \cdot L_\beta (q)
\right\}  \nonumber \\
&+& {g^2 \over 4}
\int_{-\pi}^{\pi}{d^n q\over (2\pi)^n}{\rm tr} 
\left\{
L_\alpha (q) \cdot [L_\alpha (q)\star 1 - 1 \star L_\alpha (q)] 
( L_\beta  (q) \cdot L_\alpha (q) 
      + L_\alpha (q) \cdot L_\beta  (q) )
\right\}  \nonumber \\
&+& {-g^2 \over 4}
\int_{-\pi}^{\pi}{d^n q\over (2\pi)^n}{\rm tr} 
\left\{
L_\alpha (q) \cdot L_\beta (q)
[L_\alpha (q)\star 1 - 1 \star L_\alpha (q)] \cdot L_\beta (q)
\right\},
\label{eqn:2point-b}
\end{eqnarray}
and $\Pi^{(c)}$ becomes
\begin{eqnarray}
\Pi^{(c)} &=& -\Pi^{(b)} + \nonumber \\
&- & 
{g^2 \over 8}\int_{-\pi}^{\pi}{d^n q\over (2\pi)^n} 
{\rm tr} \left\{L_\alpha (q) \cdot L_\alpha (q) \cdot
[L_\alpha (q)\star 1 - 1 \star L_\alpha (q)]\cdot L_\alpha (q)
\right\}  \nonumber \\
&+& {1 \over 3} {\rm tr} \left\{
[ (\partial_\alpha L_\alpha (q) \cdot L_\alpha (q))\star L_\alpha (q) 
\right. \nonumber \\
&& \left. \qquad 
-2(L_\alpha (q) \cdot \partial_\alpha L_\alpha (q))\star L_\alpha (q) 
 +L_\alpha (q) \star (\partial_\alpha L_\alpha (q) \cdot L_\alpha (q))]
\right\} \nonumber \\
&+& 
\mbox[H. \ C. ].
\label{2point-c}
\end{eqnarray}
\narrowtext
Here $H. \ C. $ stands for hermitian conjugate.
It can be seen easily that
$\Pi^{(b)}$ which appears in $\Pi^{(a,c)}$
satisfies the equation:
\begin{eqnarray}
\sum_t \Pi^{(b)} _{s,t} = 0.
\label{2pb}
\end{eqnarray}
A similar formula can be found for $\Pi^{(c)}$
after a simple calculation:
\begin{eqnarray}
\sum_t \Pi^{(c)} _{s,t} &=& {-g^2 \over 8}
\int_{-\pi}^{\pi}{d^n q\over (2\pi)^n}{\rm tr}  
\partial_\alpha \left\{ L_\alpha (q) ^3 \right\} = 0, 
\label{2pc}
\end{eqnarray}
since the integral of n dimensions without singularities
is zero from Stokes' theorem.

The transverse term $\Pi^{(a)}$ satisfies the following equation:
\begin{eqnarray} 
\sum_{s,t} \Pi^{(a)} &=& 
2\times {g^2 \over (4 \pi)^2 } {2 \over 3}
 \left( 
{1 \over \epsilon} + log (a^2 \bar \mu^2) + 1.147 
\right).
\label{2pa}
\end{eqnarray} 
It is noted that $\Pi^{(a)}$ has $1/\epsilon$ term.
This is an infra-red singularity, which is canceled 
in the total expression of $\Pi_{\mu\nu,st}(p)$.
This point is discussed also in Appendix B.

\

\paragraph{Total contribution of $\Pi_{\mu \nu} (p)$}\

\begin{figure}
\centerline{\epsfxsize=12cm \epsfbox{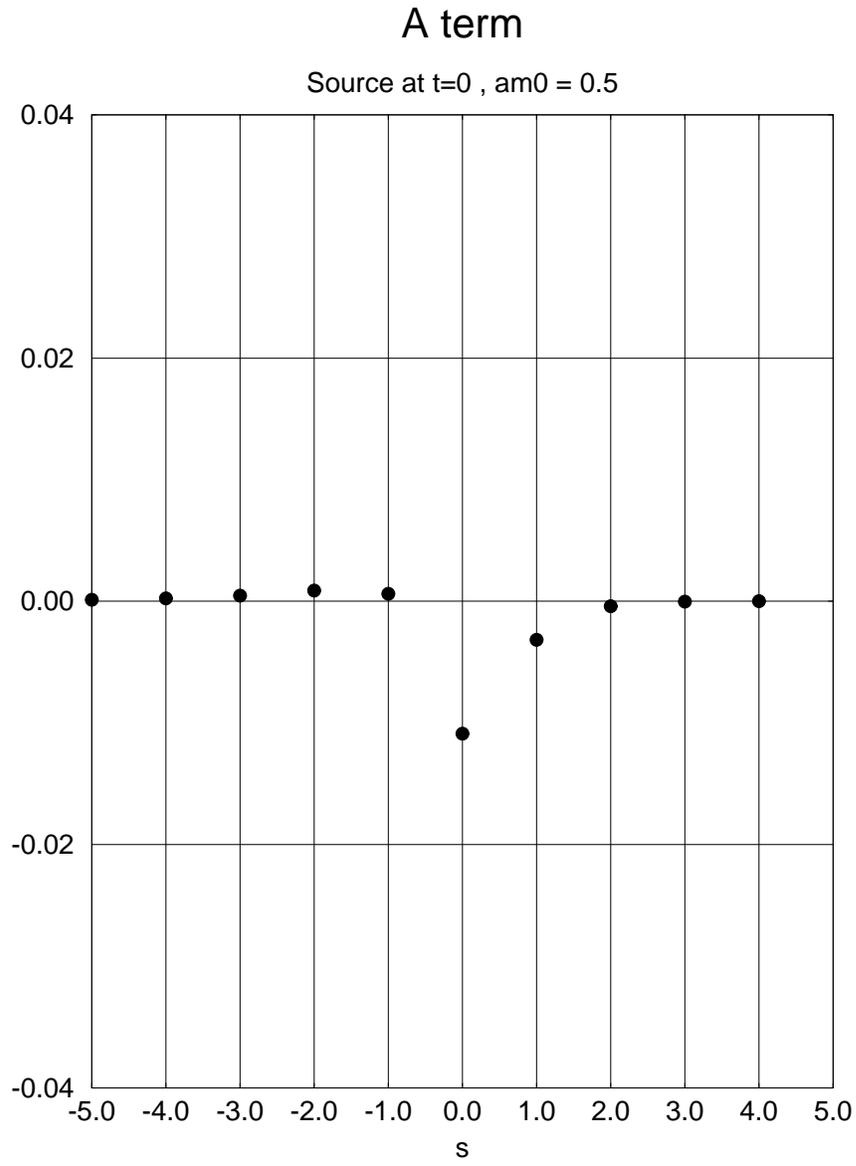}}
\caption{The coefficient of the 2 point function
$\Pi^{(a)}_{st}$ as a function of $s$ 
with $t=0$ fixed at $m_0=0.5$ and $L=5$.
Here divergent contributions are removed from $\Pi^{(a)}_{st}$.
}
\label{trans4}
\end{figure}

\begin{figure}
\centerline{\epsfxsize=12cm \epsfbox{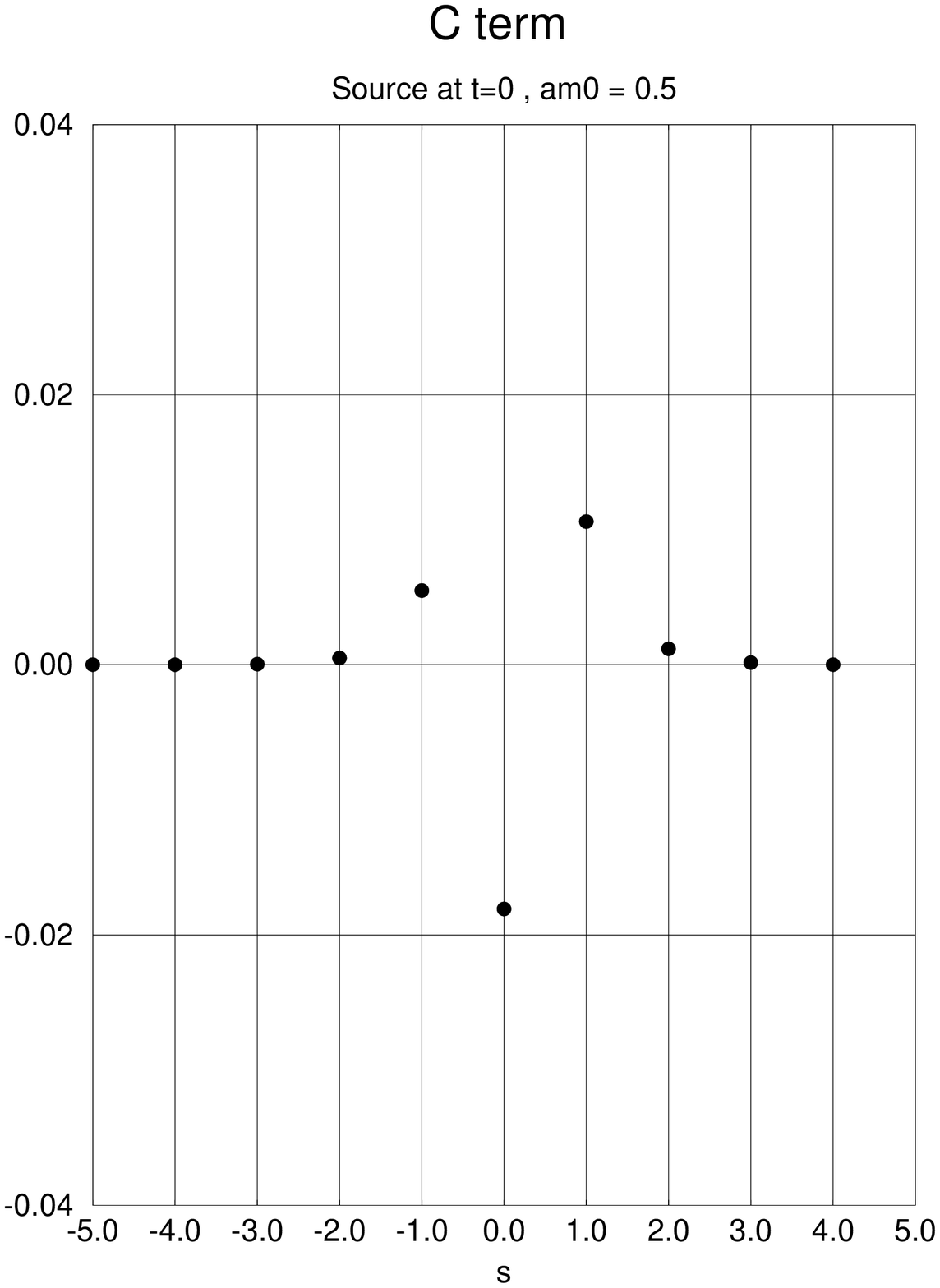}}
\caption{The coefficient of the 2 point function
$\Pi^{(c)}_{st}$ as a function of $s$ 
with $t=0$ fixed at $m_0=0.5$ and $L=5$.
}
\label{nonLo4}
\end{figure}

As $\Pi ^{cont.}_{\mu \nu} (p)$ and $\Pi ^{lattice}_{\mu \nu} (p)$ 
are obtained in the previous paragraph, the total contribution of 
$\Pi_{\mu \nu} (p)$ ($s,t$ are suppressed) is now given  by
\widetext
\begin{eqnarray}
\label{t2point}
\Pi_{\mu \nu} (p) &=&
 [ \Pi_{\mu \nu} ^{cont.}(p)]  
+[ \Pi_M
  +\Pi^{(a)} T_{\mu\nu}(p) 
  +\Pi^{(b)} \delta_{\mu\nu} p^2
  +\Pi^{(c)} \delta_{\mu\nu} p_\nu ^2] \nonumber \\
&=&
   \Pi_M
  +\Pi (p) T_{\mu\nu}(p) 
  +\Pi^{(b)} \delta_{\mu\nu} p^2
  +\Pi^{(c)} \delta_{\mu\nu} p_\nu ^2,
\end{eqnarray}
\narrowtext
where $\Pi (p)$ is the total contribution 
for the transverse term,
which has no $1/\epsilon$ term (see, Appendix B).  
Since it is difficult 
to calculate eq.(\ref{t2point}) analytically, 
we numerically evaluate each term in eq.(\ref{eqn:para}):
a finite part of $\Pi^{(a)}$ (A term), 
$\Pi ^{(b)}$ ( B term ) and $\Pi^{(c)}$ ( C term),
and plot results in Fig.\ref{trans4} $\sim$ \ref{nonLo4}.
These figures are written as a functions of $-L \le s \le L$ (L = 5)
for $t=0$ fixed.

From the above result, we draw the following properties for 
the structure of the 
gauge field 2 point function at a fermion 1 loop level, as a conclusion of this
subsection.
\begin{enumerate} 
\item 
If the gauge field has no dependence of the extra dimension,
({\sl i.e.} $A_\mu(s,n) \rightarrow A_\mu(n)$), 
the propagator $\Pi_{\mu\nu}(p)$ becomes transverse, thus gauge invariant, 
as expected.
This can be easily seen from a fact that only the transverse term 
in eq.(\ref{t2point}) is left after summation over s,t 
(using eq.(\ref{eqn:2pmass}), (\ref{2pb}), (\ref{2pc}) and (\ref{2pa}));
\begin{eqnarray}
\sum_{s,t} \Pi_{\mu\nu,s t} (p) &=&
{- g^2 \over (4\pi)^2} 2 \times {2 \over 3} T_{\mu\nu}(p) \\
&\times &
\left[{-5 \over 3}+\gamma_E 
+(\log({p^2 a^2 \over 4 \pi}) +1.147)\right].\nonumber
\end{eqnarray}

\item 
$\Pi_M$, $\Pi (p)$, $\Pi^{(b)}$ and $\Pi^{(c)}$ 
have a peak around $s=0$ for fixed $t=0$. 
This shows that there is no gauge invariance 
when the gauge field depends on the extra dimension $s,t$.
This fact will be discussed later.
\end{enumerate} 

\subsection{Anomaly in  4+1 dimensions}

The next order of the effective action is a three point function.
An important quantity which we should evaluate is a divergence of the gauge current,
{\sl i.e.} gauge anomaly.
In this subsection we shall concentrate on a calculation of the gauge anomaly.
An effective action for three gauge fields induced by a fermion loop integral
is written as
\widetext
\begin{eqnarray}
S_{eff} ^{(3)} &=& g^3
\sum_{stu}
\int_{-\pi/a}^{\pi/a} {d^n q\over (2\pi)^n}
{d^n p\over (2\pi)^n} {d^n k\over (2\pi)^n}
(2\pi)^n \delta^{(n)}(q+p+k) \nonumber \\
&& \times \Gamma_{\mu\nu\rho} (q,t,k;s,t,u) 
\times A_{\mu}(s,q)A_{\nu}(t,p)A_{\rho}(u,k),
\end{eqnarray}
where $\Gamma_{\mu\nu\rho} (q,t,k;s,t,u)$ is the three point function of the 
gauge field.

The anomaly is defined as a variation of $S_{eff}^{(3)}$ under the 
infinitesimal gauge transformation $\delta A_{\rho}(k,u) = 
\displaystyle {i \over g}  q_\mu \theta(k,u)$:
\begin{eqnarray}
\delta S_{eff} ^{(3)} &=&3 g^2
\sum_{stu}\int_{-\pi/a}^{\pi/a} {d^n q\over (2\pi)^n}
{d^n p\over (2\pi)^n} {d^n k\over (2\pi)^n}
(2\pi)^n \delta^{(n)}(q+p+k) \nonumber \\
&&  \times i k_\rho \Gamma_{\mu\nu\rho} (q,t,k;s,t,u)
    \times A_{\mu}(q,s)A_{\nu}(p,t)\theta(k,u),
\end{eqnarray}
and a relation between 
$\delta S_{eff} ^{(3)}$ and the current divergence is given by,
\begin{eqnarray}
\delta S_{eff} ^{(3)} &=& 
\sum_u \int_{-\pi/a}^{\pi/a} {d^n k\over (2\pi)^n}
\delta A_{\rho}(k,u){\delta S_{eff} ^{(3)} \over \delta A_{\rho}(k,u)}
\nonumber \\
&=& 
\sum_u \int_{-\pi/a}^{\pi/a} {d^n k\over (2\pi)^n}
\left\{ -{i \over g} \theta(k,u)  k_\mu J_\rho (k,u) \right\} .\nonumber 
\end{eqnarray}
More explicitly,
\begin{eqnarray}
k_\rho J_\rho (k,u) &=& -g^3
\sum_{st}\int_{-\pi/a}^{\pi/a} {d^n q\over (2\pi)^n}
{d^n p\over (2\pi)^n} (2\pi)^n \delta^{(n)}(q+p+k) \nonumber \\
&& \times 
 k_\rho \Gamma_{\mu\nu\rho} (q,p,k;s,t,u)
\times A_{\mu}(q,s)A_{\nu}(p,t).
\end{eqnarray}
Since  $\Gamma_{\mu\nu\rho}$ has the logarithmic divergence,
we evaluate it as follows.
\begin{eqnarray} &&
 k_\rho \Gamma_{\mu\nu\rho} (q,p,k;s,t,u)  \nonumber \\
&&= k_\rho \left[\Gamma_{\mu\nu\rho} (q,t,k;s,t,u) 
- \Gamma_{\mu\nu\rho} (0,0,0;s,t,u) \right. \nonumber \\
&&\qquad \left. 
-q_\sigma \partial_\sigma \Gamma_{\mu\nu\rho} (0,0,0;s,t,u)
-p_\sigma \partial_\sigma \Gamma_{\mu\nu\rho} (0,0,0;s,t,u)
\right] \nonumber \\
&& + k_\rho 
\left[\Gamma_{\mu\nu\rho} (0,0,0;s,t,u) \right. \nonumber \\
&& \qquad \qquad \left.
+q_\sigma \partial_\sigma \Gamma_{\mu\nu\rho} (0,0,0;s,t,u)
+p_\sigma \partial_\sigma \Gamma_{\mu\nu\rho} (0,0,0;s,t,u)
\right]. \nonumber
\end{eqnarray}
We can then evaluate the first term using the continuum form
as in the previous subsection.
Only the parity odd part of $\Gamma_{\mu \nu \rho}$ is necessary
to get the anomaly, and it becomes
\begin{eqnarray}
 k_\rho \Gamma_{\mu\nu\rho} (q,p,k;s,t,u)^{odd} &=&
 k_\rho \left[\Gamma_{\mu\nu\rho} (q,t,k;s,t,u) 
\right]^{cont.} \nonumber \\
&+& k_\rho \left[
q_\sigma \partial_\sigma \Gamma_{\mu\nu\rho} (0,0,0;s,t,u) \right.
\nonumber \\ 
&& \qquad
\left.
+p_\sigma \partial_\sigma \Gamma_{\mu\nu\rho} (0,0,0;s,t,u)
\right]^{lattice}.
\label{3poreg}
\end{eqnarray}
\narrowtext
Here we use a fact that the anomaly has an anti-symmetric structure for
Lorentz indices, $\mu,\nu,\rho,\sigma$.

We now perform the calculation of these terms, 
continuum and lattice ones, in the next two paragraphs, and the detail is
given in Appendix C.

\

\paragraph{Evaluation for the Continuum part}\

Althouh one can use the dimensional regularization as in the previous 
subsection to calculate the continuum part of the anomaly,
we introduce a different regularization here
---$\displaystyle \int_{-\infty} ^\infty$ 
is replaced by $\displaystyle \int_{-\Lambda} ^\Lambda$ in $n=4$ ($
\displaystyle \Lambda = {\pi \over a}$).
Accordingly, using 
$F_\chi ^3  = F_\chi (s,t) F_\chi (t,u) F_\chi (u,s)$,
we can write
\widetext
\begin{eqnarray}
&&k_\rho \left[\Gamma_{\mu\nu\rho} (q,p,k;s,t,u)\right]^{cont.} 
\nonumber \\
&=&  {1 \over 3}
g^3 \sum_\chi F_\chi ^3
\int_{-\Lambda}^\Lambda {d^4 l\over (2\pi)^4} 
{\rm tr} \left\{
i \gamma_\mu P_\chi 
{-i {l\!\!\!/\,\,} \over l^2}
i \gamma_\nu P_\chi 
{-i ({l\!\!\!/\,\,}+{p\!\!\!/\,\,}) \over (l+p)^2}
i {k\!\!\!\!/\,\,}P_\chi 
{-i ({l\!\!\!/\,\,}-{q\!\!\!/\,\,}) \over (l-q)^2}
\right\} \nonumber \\
&&
\qquad \qquad + [(l \rightarrow l-p)+ (l \rightarrow l-q)]
\nonumber \\
&=&  {1 \over 3}
g^3 \sum_\chi F_\chi ^3
\int_{-\Lambda}^\Lambda {d^4 l\over (2\pi)^4} 
{\delta_\chi \over 2 } 4 i \epsilon_{\mu\nu\rho\sigma}
\left({l^\rho (l+p)^\sigma \over l^2 (l+p)^2}
     - {l^\rho (l-q)^\sigma \over l^2 (l-q)^2}\right) 
+[(l \rightarrow l-p)+ (l \rightarrow l-q)]. \nonumber 
\end{eqnarray}
Here we use the Ward-Takahashi identity as usual in the last step.
It is noted that we can not make this integral zero 
using the shift $l \rightarrow l+q$ etc,
since the integral with respect to $l$ has boundaries,
$-\Lambda$ or $\Lambda$.

Using the following integration formula:
\begin{eqnarray}
\int_0 ^1 dx 
\int_{-\Lambda -x p} ^{\Lambda -x p} {d^4 l \over (2\pi)^4} 
{\partial \over \partial l_\alpha} {1 \over l^2 +h^2(x)}
&=& {1 \over 16 \pi^2},(\Lambda \rightarrow \infty),
\end{eqnarray}
where $h^2(x)$ is a positive function of$ x$,
we finally obtain 
\begin{eqnarray}
\left[k_\rho \Gamma_{\mu\nu\rho} (q,p,k;s,t,u)\right]^{cont.} 
&=& {i g^3 \over 24 \pi^2}
\sum_\chi \delta_\chi F^3 _\chi
\epsilon_{\mu\nu\rho\sigma} q_\rho p_\sigma .
\end{eqnarray}
\

\narrowtext
\paragraph{Evaluation for the Lattice part}\

The lattice part of the anomaly is written as
\begin{eqnarray}
k_\rho \left[
q_\sigma \partial_\sigma \Gamma_{\mu\nu\rho} 
+p_\sigma \partial_\sigma \Gamma_{\mu\nu\rho}
\right]^{lattice}  (0,0,0;s,t,u) \nonumber \\
\qquad \qquad \qquad \qquad  
= \epsilon_{\mu\nu\rho\sigma} q_\rho p_\sigma A_{q,p}(s,t,u),
\label{anolat1}
\end{eqnarray}
where  $A_{q,p}(s,t,u) = A_q(s,t,u)-A_p(s,t,u)$ and
\begin{eqnarray}
A_q(s,t,u) 
= {1 \over 4!} \epsilon_{\mu\nu\rho\sigma} 
{\partial \over \partial q_\sigma}
 \Gamma_{\mu\nu\rho} (0,0,0;s,t,u). \nonumber 
\end{eqnarray}
An explicit form of $A_{q,p}(s,t,u)$ is given as
\widetext
\begin{eqnarray}
A_{q,p}(s,t,u) &=& 
{1 \over 4!} g^3 \epsilon_{\mu\nu\rho\sigma} {1 \over 3}
\int_{-\pi}^\pi {d^4 l\over (2\pi)^4} 
3 {\rm tr} \left\{ 
\left(L_\mu (l) \cdot L_\sigma (l)\right) 
\star L_\nu (l) \star L_\rho (l)
\right\} \nonumber \\
&&\qquad \qquad \qquad \qquad \qquad -\partial_\sigma
{\rm tr} \left\{ 
L_\mu (l) \star L_\nu (l) \star L_\rho (l)
\right\}
\label{anolat}
\end{eqnarray}
\narrowtext
where the extra dimension indices $s,t,u$ are implicit
in the above equation and
$[L_\mu]_{st} (l)
=[\partial_\mu S_F ^{-1}(l)\cdot S_F(l)]_{st}$
is used again.

As seen in the calculation of the two point function,  
there are two useful formulae:
\begin{eqnarray}
\sum_u A_{q,p}(s,t,u) = -{i g^3 \over 24 \pi^2}
\sum_\chi \delta_\chi F^2_\chi (s,t)  
\epsilon_{\mu\nu\rho\sigma} q_\rho p_\sigma 
\label{3pou}
\end{eqnarray}
and
\begin{eqnarray}
\sum_{st} A_{q,p}(s,t,u) &= &
{i g^3 \over 12 \pi^2}
\sum_\chi \delta_\chi F_\chi (u,u)  
\epsilon_{\mu\nu\rho\sigma} q_\rho p_\sigma .
\label{3post}
\end{eqnarray}
The detailed proof is given also in Appendix C.

\paragraph{Results}\

A total contribution of the parity-odd term, the gauge anomaly, becomes
\begin{eqnarray}
k_\rho \Gamma_{\mu\nu\rho} (q,p,k;s,t,u)^{odd}= 
\epsilon_{\mu\nu\rho\sigma} q_\rho p_\sigma  &&
\nonumber 
\end{eqnarray}
\begin{equation}
\qquad \qquad \qquad \times \left[
{i g^3 \over 24 \pi^2}
\sum_\chi \delta_\chi F^3 _\chi (s,t,u) 
+ A_{q,p} (s,t,u)  \right].
\end{equation}

\begin{figure}
\centerline{\epsfxsize=12cm \epsfbox{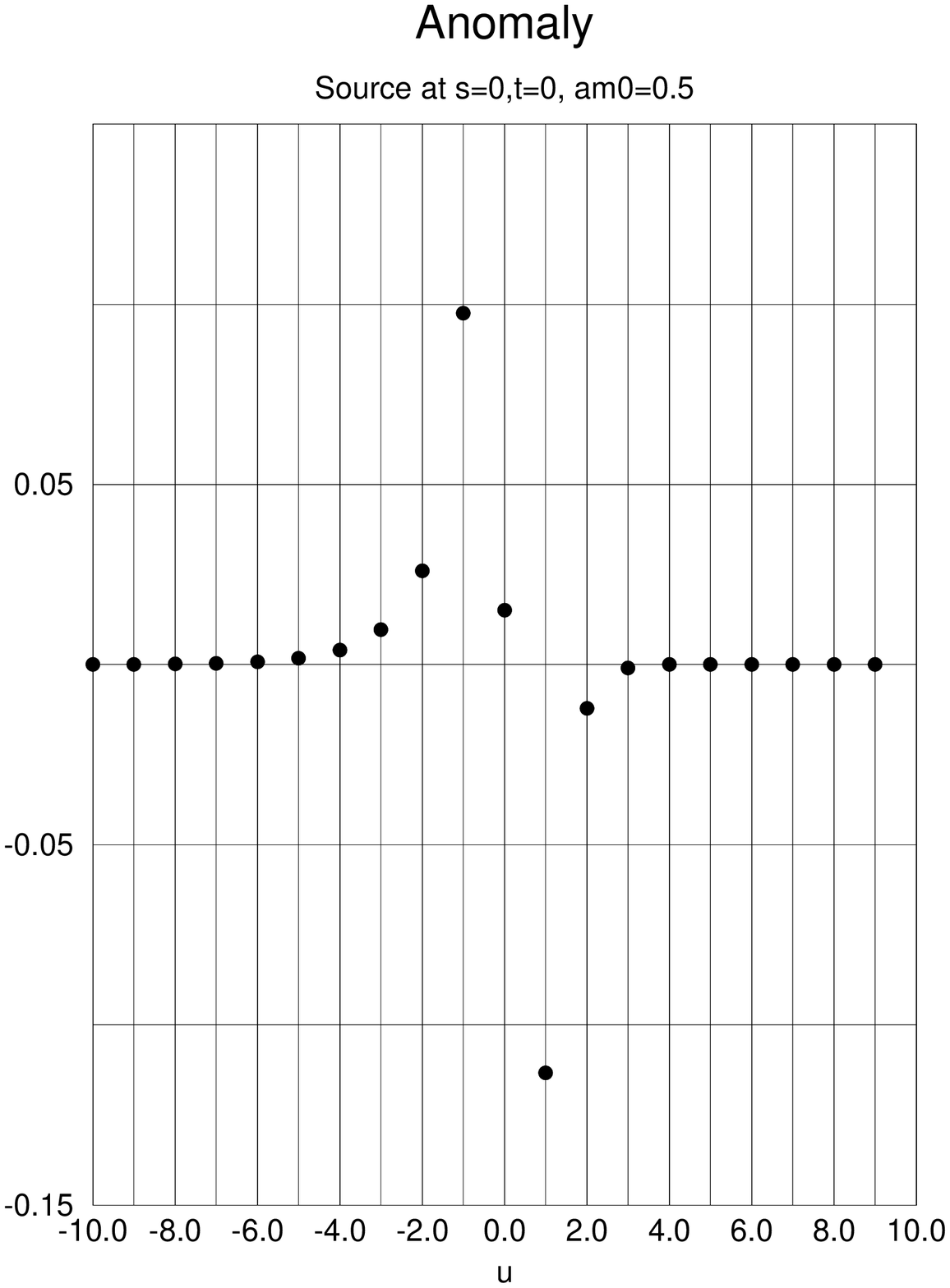}}
\caption{The coefficient of the anomaly $C(s,t,u)$ in 4+1 dimensions
as a function of $u$ for  $s,t=0$ fixed
at $m_0=0.5$ and $L=10$.
}
\label{anomaly4st_1}
\end{figure}

\begin{figure}
\centerline{\epsfxsize=12cm \epsfbox{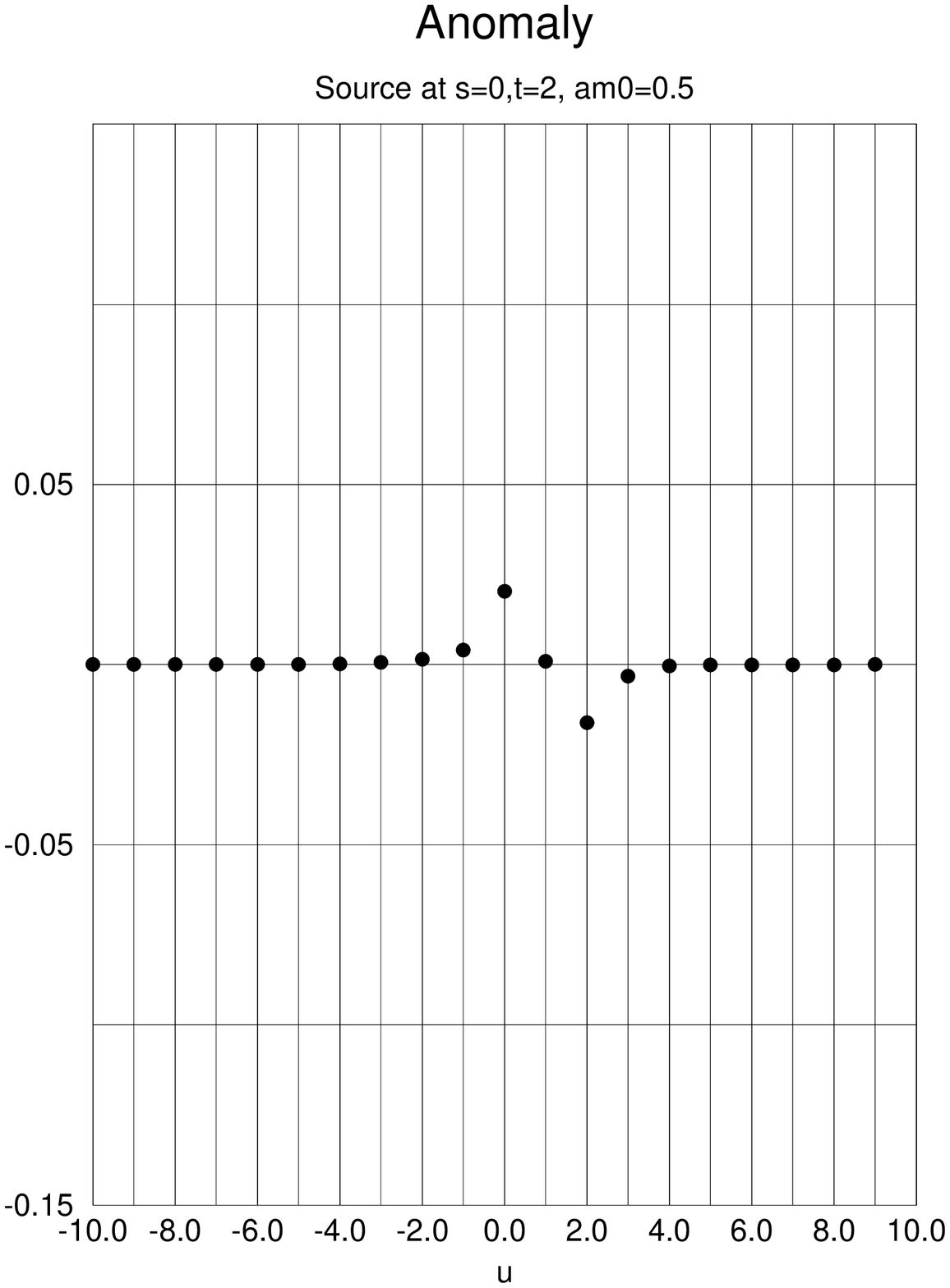}}
\caption{The coefficient of the anomaly $C(s,t,u)$ in 4+1 dimensions
as a function of $u$ for  $s=0$ and $t=2$ fixed
at $m_0=0.5$ and $L=10$.
}
\label{anomaly4st_2}
\end{figure}

\begin{figure}
\centerline{\epsfxsize=12cm \epsfbox{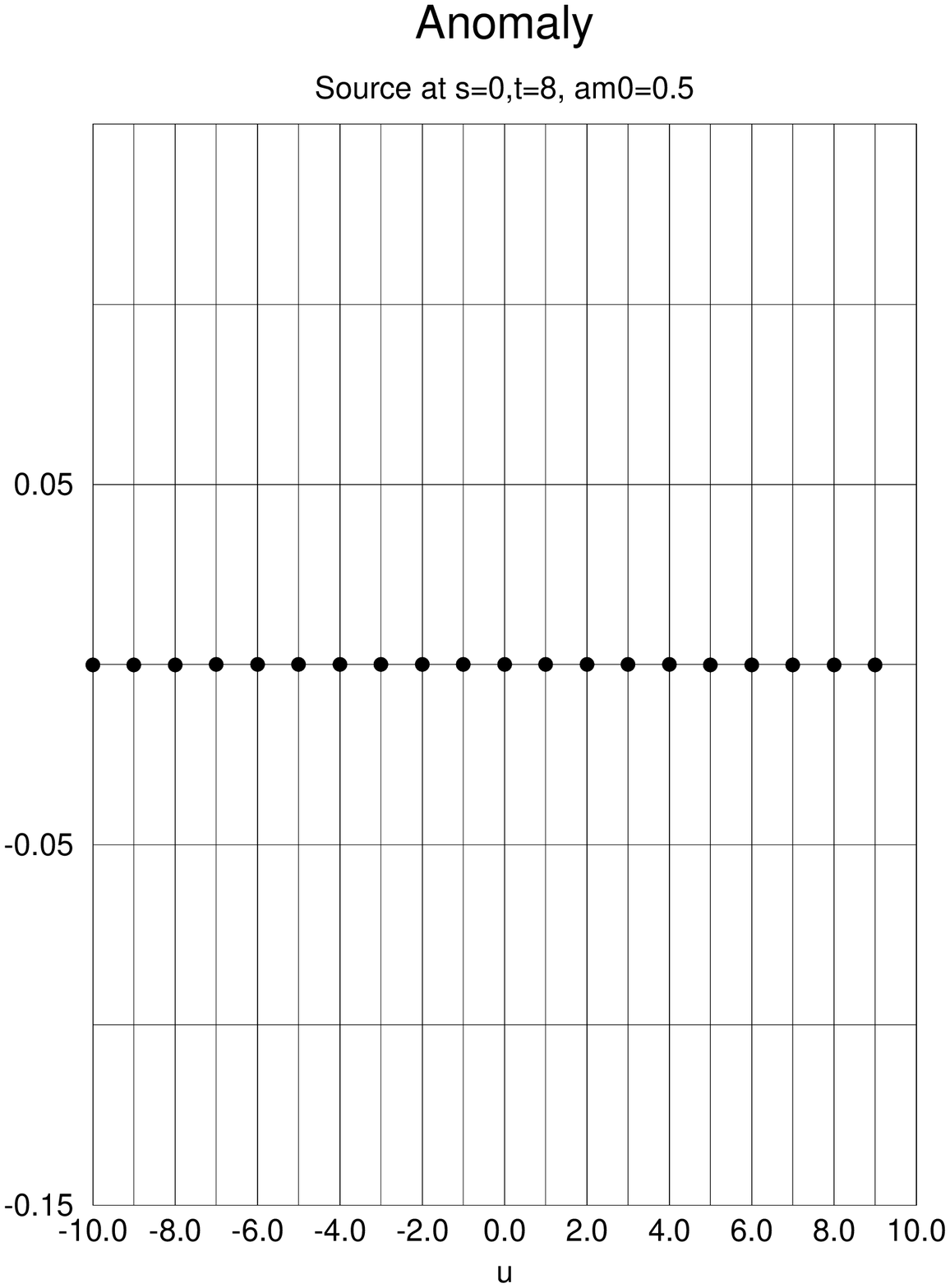}}
\caption{The coefficient of the anomaly $C(s,t,u)$ in 4+1 dimensions
as a function of $u$ for  $s=0$ and $t=8$ fixed
at $m_0=0.5$ and $L=10$.
}
\label{anomaly4st_3}
\end{figure}

\begin{figure}
\centerline{\epsfxsize=12cm \epsfbox{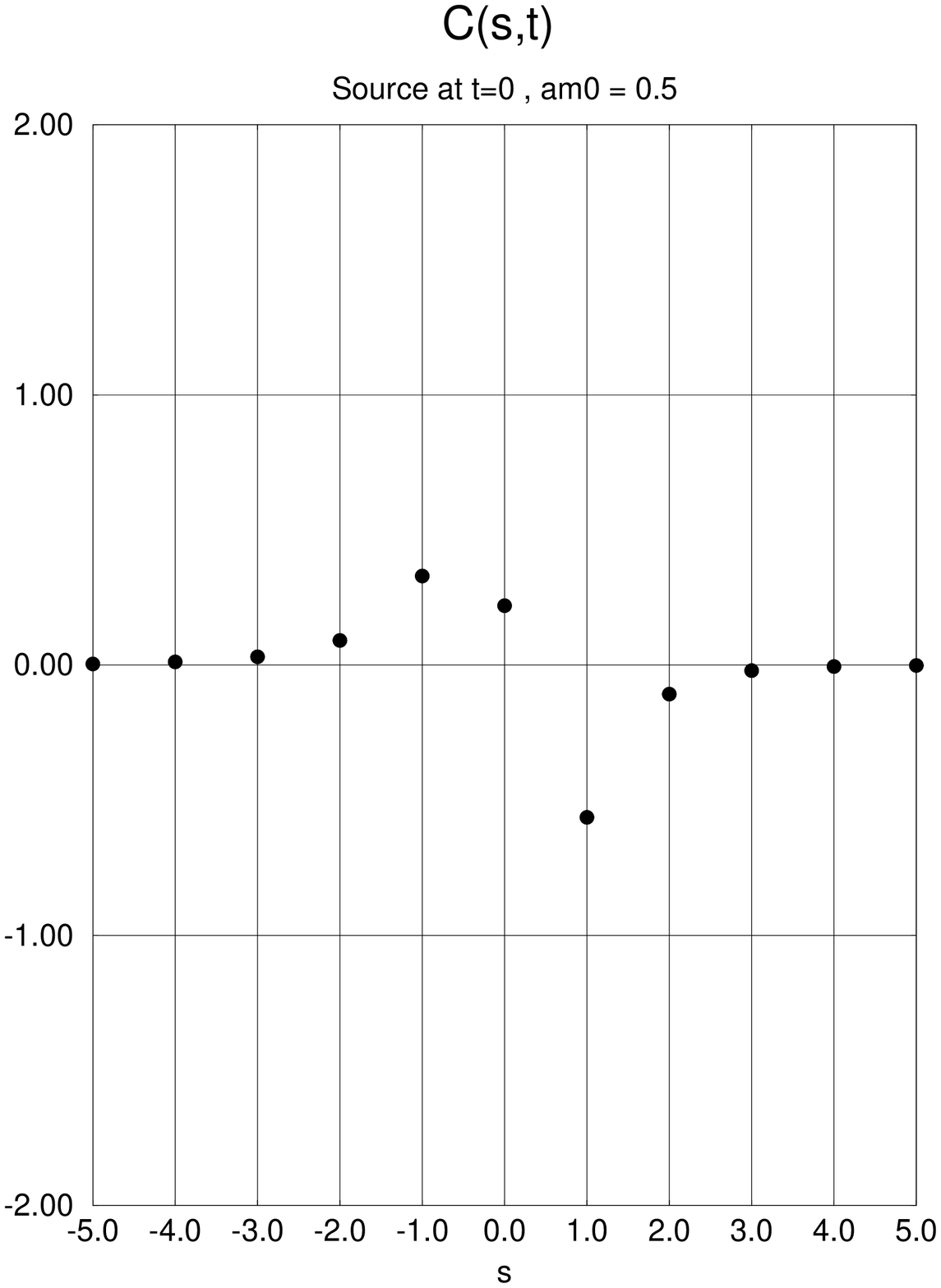}}
\caption{The coefficient of the anomaly $C(s,t)$ in 2+1 dimensions
as a function of $s$ for  $t=0$ fixed
at $m_0=0.5$ and $L=5$.
}
\label{anomaly2s}
\end{figure}

We summarize important properties of the anomaly as follows.
\begin{enumerate}
\item
Note that 
a summation over u makes this anomaly zero due to eq.(\ref{3pou}).
This comes from the fact
that the model with the axial gauge fixing is invariant 
under $s$ independent gauge transformations.

\item
Because of eq.(\ref{3post}),
a summation over s,t makes this anomaly equal to
\widetext
\begin{eqnarray}
\sum_{st} k_\rho \Gamma_{\mu\nu\rho} (q,p,k;s,t,u)|_{anomaly} &=&
{i g^3 \over 8 \pi^2}
\epsilon_{\mu\nu\rho\sigma} q_\rho p_\sigma
\sum_\chi \delta_\chi F_\chi (u,u). 
\label{eqn:anom_sum}
\end{eqnarray}
\narrowtext
Physically this is the gauge anomaly at $u$ for $s$ independent
background gauge field $A_\mu (s,n) = A_\mu (n)$.
It is noted that the coefficient of eq.(\ref{eqn:anom_sum}), 
$\displaystyle {1\over 8 \pi^2}$, is equal to the coefficient of the
covariant anomaly in 4 dimensions, not the one of the consistent 
anomaly in 4 dimensions. This does not contradict with the fact that
the above anomaly is derived from the variation of the effective action,
since the effective action is defined in 4 + 1 dimensions, not in 4 dimensions.

\item
Without summations
it is difficult to calculate the anomaly analytically.
Instead we evaluate it numerically, and 
results are given in Figs.\ref{anomaly4st_1}--\ref{anomaly4st_3}.
We plot
$ [\displaystyle {i g^3 \over 24 \pi^2}
\sum_\chi \delta_\chi F^3 _\chi (s,t,u) 
+ A_{q,p} (s,t,u)  ]$ as a function of $u$ 
at $m_0 =0.5$ and $L=10$,
for fixed $s=t=0$ in Fig.\ref{anomaly4st_1},
for $s=0$ and $t=2$ in Fig.\ref{anomaly4st_2}, and
for $s=0$ and $t=8$ in Fig.\ref{anomaly4st_3}.
These figures tell us the following.
If two gauge fields $A_\mu (s)$ and $A_\nu (t)$ are on the same 4 dimensional
subspace ($s=t=0$), the anomaly has large contribution around $u=s=t$.
If $t$ differs a little from $s$ ( $s=0$ and $t=2$ ), the anomaly has non-zero
contribution around $u=s$ or $u=t$ but peak heights become less.
If $t$ is far away from $s$ ($s=0$ and $t=8$ ), the anomaly almost vanishes
at all $u$. For comparison we calculate the anomaly in 2+1 dimensions, which is
given by\cite{Our}
\begin{equation}
T_{\rm anomaly}(s,x)=\sum_t C(s,t) {i g^2\over 4\pi}\epsilon^{\mu\nu}
\delta_\mu A_\nu (t,x)
\end{equation}
where
\begin{equation}
C(s,t)=\sum_X\delta_X F_X(s,t)^2-2\Gamma_{CS}(s,t).
\end{equation}
In Fig.\ref{anomaly2s} $C(s,t)$ is plotted as a function of $s$ with $t=0$ 
fixed at $m_0$=0.5 and $L=5$.
Behaviors of the gauge anomalies are similar both in 4+1 dimensions and in
2+1 dimensions: they have large contribution around $s=0$ or $u=0$.
\end{enumerate}

Finally it should be mentioned that a parity-even part of the variation
of the 3 point function under the gauge transformation vanishes locally
for $s$ independent background gauge fields.

\subsection{The 4-point function}


A calculation of a 4-point function 
is simpler than the previous ones since there is no derivative term.
Using the same logic as before we obtain
\widetext
\begin{eqnarray}
\Gamma_{\mu\nu\rho\sigma} ^{(4)}(q,p,k,r;s,t,u,v)
&&=
\left[
\Gamma_{\mu\nu\rho\sigma} ^{(4)}(q,p,k,r;s,t,u,v) 
\right]^{cont.} 
+\left[
\Gamma_{\mu\nu\rho\sigma} ^{(4)}(0,0,0,0;s,t,u,v)  
\right]^{lattice}.
\nonumber
\end{eqnarray}
\narrowtext

Results are very similar to the 2- or 3-point functions;
After the summation over $v$
the gauge variation of the 4-point function is zero 
and after the summation over $s,t,u $
it reproduces the continuum value.
It shows that
the gauge invariance is correct only 
in the case 
that the gauge field has no dependence on the extra dimension.

\section{Physical Implication}
\label{sec:PI}

We consider the 4+1 dimensional theory so far.
Properties of the effective action in 4+1 dimensions are 
more or less similar to those in 2+1 dimensions 
except that the 2-point function of the gauge fields contains
a mass term and a Lorentz non-covariant term 
as well as a transverse term and a longitudinal term,
and that the gauge anomaly is proportional to the charge cubed.
As in the 2 dimensional case\cite{Our} 
the fermion number violation can be incorporated as follows.
First let us consider a fermion number current, whose  expectation value
for background gauge fields is defined by
\begin{equation}
\langle J_\mu^g (s,x) \rangle = \frac{\delta S_{eff}^{(3,odd)}}{g\delta
A_\mu (s,x)}
\end{equation}
where the index $g$ in the current explicitly shows the charge of the fermion,
and the gauge current  is equal to $ g J_\mu^g (s,x) $.
A divergence of the fermion number current, the fermion number
anomaly, is proportional to $g^2$ while the gauge anomaly is proportional 
to $g^3$. Therefore even for anomaly-free model such as the standard model,
where a sum of charge cubed vanishes, the fermion number violation, 
which is proportional to a sum of charge squared, can remain non-zero 
in the domain-wall fermion formulation.
This is the most important consequence of our results:
The domain-wall fermion formulation may allow us to simulate 
the non-perturbative dynamics of the baryon number violation
of the standard electro-weak theory.
Although our results are obtained for a U(1) gauge theory, it is not so 
difficult to obtain similar results for non-abelian gauge models. 

The gauge non-invariant terms remain non-zero in 4(+1) dimensions 
even for anomaly-free cases as in 2(+1) dimensions.
In the next section we will briefly mention a way to avoid these terms.
The $s$-independent gauge fields again lead to the vector-like theory
as in the 2 dimensional case\cite{Our}.

\section{Conclusion}
\label{sec:Con}

In this paper we have investigated
the domain-wall chiral fermion formulation in 4+1 dimensions
with a lattice perturbation theory. 
We have calculated in detail 
the 2-point function and the gauge anomaly in 4 dimensional
U(1) chiral gauge theory formulated in 4+1 dimensions.

The most important conclusion 
drawn from our perturbative analysis is that
the baryon number violation of the standard model may be incorporated 
by the domain-wall fermion formulation.

However, as stressed in the previous section, 
the domain-wall fermion formulation can not maintain the transversality,
and hence gauge invariance,
even for anomaly-free cases.

A solution to the gauge invariance has been proposed by 
Narayanan and Neuberger:
They take the size of the extra dimension strictly infinite,
instead of periodic box, 
and make gauge fields independent of the extra dimension.
(They also take the ``temporal gauge'' from the beginning.)
Since no anti-domain wall caused previously by the periodicity
of the extra dimension exists, 
the theory is expected to remain chiral even for such gauge fields.
Indeed, using our result of the effective action in 2+1 dimensions 
it has been shown that the Narayanan-Neuberger formulation, called
``overlap formula'', 
gives gauge invariant effective theory 
except for the gauge anomaly in 2+1 dimensions\cite{AL}.
The U(1) chiral gauge theory via their formulation in 4+1 dimensions 
is now under investigation
since our calculation in this paper can be easily extended to it.
As mentioned in the introduction, however, it is pointed out recently 
that the overlap formula can not reproduce the desired chiral zero mode
in the presence of the rough gauge dynamics ( gauge degree of freedom )
which appears due to the violation of the gauge invariance at infinities. 
This problem is known to exist also for the wave-guide model.

Except the problem of the gauge invariance above, the domain-wall fermion
formulation in 4+1 dimensions works well for smooth back-ground gauge fields. 
A remaining question of the domain-wall fermion formulation is
whether the chiral zero mode can survive 
in the presence of the dynamical gauge fields.
A numerical investigation has been performed in 2+1 dimensions but
has failed to give a definite conclusion on an existence of the zero mode
in the symmetric phase\cite{AN}.
Further numerical study for a 4+1 dimensional model is needed.

\section*{Acknowledgements}

We would like to thank Prof. A. Ukawa for helpful suggestions.
Numerical integrations for the present work have been carried out
at Center for Computational Physics, University of Tsukuba.
This work is supported in part by the Grants-in-Aid of the Ministry of
Education(Nos. 04NP0701, 06740199).

\section*{Appendix A}
\label{sec:AppA}

In this appendix, we calculate the function $G_{L/R}(p)$ .
The fundamental equations are 
\begin{eqnarray}
\bar p^2 G_L (p)_{s,t} + (M^\dagger M G_L (p))_{s,t} &=& \delta_{s,t},
\nonumber \\
\bar p^2 G_R (p)_{s,t} + (M M^\dagger G_R (p))_{s,t}&=& \delta_{s,t},
\label{gap}
\end{eqnarray} 
where $\sum_{\mu = 1} ^{2k} \sin^2(p_\mu) $ 
is written as $\bar p^2 $.
We examine the upper equation in detail. 
Explict form of this equation is
\widetext
\begin{eqnarray}
(\bar p^2 + A(s)) G_{s,t}(p) 
-B(s+1) G (p)_{s+1,t} - B(s)G (p)_{s-1,t} = \delta_{s,t},
\label{gleq}
\end{eqnarray}
\narrowtext
where
\begin{eqnarray}
A(s)&=&1+B(s)^2, \\
B(s)&=&a(s)-{r \over 2} F(p) \nonumber \\
&=&1-m(s)-{r \over 2} \sum_\mu (\cos(p_\mu)-1).
\end{eqnarray}
Since $m(s)$ in $B(s)$ changes the value at $s=0,L$,
eq.(\ref{gleq}) should be separated into two cases,  
one for $0 \le s \le L-1$ 
and the other for $-L-1 \le s \le -1$.
For convenience, $G(p)^+$ is defined as 
$G(p)$ in the range of $0 \le s \le L-1$
and $G(p)^-$ in $-L-1 \le s \le -1$.

We first focus our attention on $G^+ (p)$.
In the range $0 \le s \le L-1$,
using $B(s)=B(s+1)= 1-m_0-{r \over 2} F(p) = a_+$,
eq.(\ref{gleq}) is rewritten as follows:
\widetext
\begin{eqnarray}
(\bar p^2 + 1+a_+ ^2) G_{s,t} ^+ (p) 
-a_+ (G (p)_{s+1,t} ^+ + G (p)_{s-1,t} ^+ ) = \delta_{s,t}.
\end{eqnarray}

\narrowtext
The solution of this equation 
is expressed as a sum of homogeneous general solutions and 
an inhomogeneous special one.
Homogeneous general solutions 
with two unknown functions $g_+ ^{(1)}(t)$ and $ g_+ ^{(2)}(t)$ are
\begin{eqnarray}
g_+^{(1)}(t) e^{-\alpha_+ (p) s} + g_+^{(2)}(t) e^{\alpha_+ (p) s},
\end{eqnarray}
where $\cosh (\alpha_+ (p)) = \displaystyle
{1 \over 2}(a_+ + {1+ \bar p^2 \over a_+})$.

Next the inhomogeneous solution is calculated 
with the Fourier transformation as
($l_n = 2 n \pi  / L $)
\widetext
\begin{eqnarray}
{1 \over 2 L a_+ } \sum_{n=0}^{L-1} 
{e^{i l_n |s-t|} \over \cosh(\alpha_+ (p)) - \cos (l_n)}
&=& 
{-1 \over 2 a_+ L \sinh (\alpha_+(p)) } 
\sum_{n=0}^{L-1} [{Z^{|s-t|+1} \over Z - e^{ \alpha_+ (p)}}
                 -{Z^{|s-t|+1} \over Z - e^{-\alpha_+ (p)}}] \nonumber \\
&=&  {\cosh(\alpha_+ (p) (|s-t|- L/2)) 
\over 2 a_+ L \sinh (\alpha_+(p)) \sinh (\alpha_+(p) L)},
\end{eqnarray}
\narrowtext
where the following formula is used in the last step:
\begin{eqnarray}
{1 \over L } \sum_{n=0}^{L-1} {Z^{s+1} \over Z - \zeta}
= {\zeta ^s \over 1 - \zeta ^L }, \qquad (Z = e^{i l_n }).
\end{eqnarray}
Therefore, the solution $G(p)_{s,t}^+ $ is given as
\begin{eqnarray}
G(p)_{s,t}^+ 
&=& g_+^{(1)}(t) e^{-\alpha_+ (p) s}
  + g_+^{(2)}(t) e^{ \alpha_+ (p) s} \nonumber \\
&+&   {\cosh(\alpha_+ (p) (|s-t|- L))
 \over 2 a_+ L \sinh (\alpha_+(p)) \sinh (\alpha_+(p) L)}.
\end{eqnarray}
In same way, the solution $G(p)_{s,t}^- $ is obtained as
\begin{eqnarray}
G(p)_{s,t}^- 
&=& g_-^{(1)}(t) e^{ \alpha_- (p) s} 
  + g_-^{(2)}(t) e^{-\alpha_- (p) s} \nonumber \\
&+&   {\cosh(\alpha_- (p) (|s-t|-L))
 \over 2 a_- L \sinh (\alpha_-(p)) \sinh (\alpha_-(p) L)},
\end{eqnarray}
where 
$a_- = 1+ m_0 -{r \over 2} F(p),
\cosh (\alpha_- (p)) =\displaystyle
 {1 \over 2}(a_- + {1+ \bar p^2 \over a_-}).$

The unknown functions $g_\pm ^{(1)} (t)$ and $g_\pm ^{(2)}(t)$
are determined by the four boundary conditions,
which are obtained by considering eq.(\ref{gleq}) at $s=0$ and $L$:
\begin{eqnarray}
G(p)_{s=-1,t}^+ &=& G(p)_{s=-1,t}^-, \nonumber \\
a_+ G(p)_{s=0,t}^+&=&a_- G(p)_{s=0,t}^-, \nonumber \\
G(p)_{s=L-1,t}^+&=& G(p)_{s=-L-1,t}^-, \nonumber \\
a_+ G(p)_{s=L,t}^+&=&a_- G(p)_{s=-L,t}^-.
\end{eqnarray}
Explicitly,
\begin{eqnarray}
 \Theta \cdot \vec{g}(t) = \vec{b}(t), \label{glprop} 
\end{eqnarray}
\widetext
\begin{eqnarray}
\Theta &=&
\left(
\begin{array}{cccc}
 e^{\alpha_+} & e^{-\alpha_+} & -e^{-\alpha_-}& -e^{-\alpha_-}\\
 a_+          & a_+           & -a_-          & -a_-          \\ 
 a_+ e^{-\alpha_+ L}         &   a_+ e^{\alpha_+ L}  &
-a_- e^{-\alpha_- L}         &  -a_- e^{\alpha_- L}    \\
     e^{-\alpha_+ (L-1)}     &   e^{\alpha_+ (L-1)}  & 
    -e^{-\alpha_- (L+1)}     &  -e^{\alpha_- (L+1)}    \\
\end{array} 
\right), \quad
\vec{g}(t)
=
\left(
\begin{array} {c}
g_+^{(1)} (t)\\
g_+^{(2)} (t)\\
g_-^{(1)} (t)\\
g_-^{(2)} (t)\\
\end{array} 
\right), \nonumber \\
\vec{b}(t) &=&
\left(
\begin{array} {l}
   X_- \cosh(\alpha_- (|1+ t|-L))   \\
\qquad \qquad \qquad      -    X_+ \cosh(\alpha_+ (|1+t|-L))\\
a_-X_- \cosh(\alpha_- (|   t|-L))   \\
\qquad \qquad \qquad -a_+ X_+ \cosh(\alpha_+ (|  t|-L))\\
a_-X_- \cosh(\alpha_- (|-L-t|-L))   \\
\qquad \qquad \qquad -a_+ X_+ \cosh(\alpha_+ (|L-t|-L))\\
   X_- \cosh(\alpha_- (|-L-1-t|-L)) \\
\qquad \qquad \qquad - X_+ \cosh(\alpha_+ (|L-1-t|-L))\\
\end{array} 
\right),\nonumber
\end{eqnarray}
where $X_\pm ^{-1} = 2 a_\pm \sinh(\alpha_\pm )\sinh(\alpha_\pm L ).$
Although solution to these equations is very complicated in general,
it becomes simpler in the limit of $ L \rightarrow \infty $.
For $t \simeq 0$,
\begin{eqnarray}
\Theta & \rightarrow &
\left(
\begin{array}{cccc}
 e^{\alpha_+} & e^{-\alpha_+} & -e^{-\alpha_-}& -e^{-\alpha_-}\\
 a_+          & a_+           & -a_-          & -a_-          \\ 
 0            & a_+ e^{\alpha_+ L}  & 0       &  -a_- e^{\alpha_- L} \\
 0            &  e^{\alpha_+ (L-1)} & 0       &  -e^{\alpha_- (L+1)} \\
\end{array} 
\right),\nonumber \\
\vec{b}(t) & \rightarrow &
\left(
\begin{array} {c}
   X_- \cosh(\alpha_- (|1+ t|-L))   -    X_+ \cosh(\alpha_+ (|1+t|-L))\\
a_-X_- \cosh(\alpha_- (|   t|-L))   -a_+ X_+ \cosh(\alpha_+ (|  t|-L))\\
 0 \\
 0 \\
\end{array} 
\right). \nonumber
\end{eqnarray}
\narrowtext
Since $g_\pm^{(2)} \rightarrow 0 $ in the limit of $L \rightarrow \infty$,
$g_\pm^{(1)}$ is easily obtained from this matrix. 
It is noted that at $t \simeq L$,  $g_\pm^{(1)} \rightarrow 0 $ in 
the limit of $L \rightarrow \infty$.
This shows that 
$G_L(p)$ has contributions from two chiral zero modes
which live at $s = 0$ or $L$.

\section*{Appendix B}
\label{AppB}

In this Appendix B, we show in detail
the calculation of the 2-point function $\Pi_{\mu\nu} (p)$
in 4+1 dimensions.

Let us set $\mu = \alpha$ and $\nu = \beta$ 
($\alpha \ne \beta $) in the parametrization of eq.(\ref{eqn:para}). 
The extra dimension indices $s,t$ are suppressed below.
For this choice of $\mu$ and $\nu$ eq.(\ref{eqn:para})  becomes
\begin{eqnarray}  
{p_\rho p_\sigma \over 2} 
\partial_{\rho,\sigma} ^2 \Pi_{\alpha \beta} ^{lattice}(0) 
&=& \Pi^{(a)} (-p_\alpha p_\beta). \nonumber 
\end{eqnarray}  
It means that
\begin{eqnarray}  
\Pi^{(a)} &=& 
 -\partial_{\alpha,\beta}^2 \Pi_{\alpha \beta} ^{lattice}(0). 
\label{pa}
\end{eqnarray}  
(No sum over $\alpha$ and $\beta$.)
When we set $\mu = \nu = \alpha$, eq.(\ref{eqn:para}) becomes
\widetext
\begin{eqnarray}  
{p_\rho p_\sigma \over 2} 
\partial_{\rho,\sigma} ^2 \Pi_{\alpha \alpha} ^{lattice}(0) 
&=& (\Pi^{(a)} + \Pi^{(b)} ) (p^2 -p_\alpha ^2 )
+(\Pi^{(b)} + \Pi^{(c)} ) p_\alpha ^2 .
\end{eqnarray}  
\narrowtext
Therefore relations between $\Pi^{(b),(c)}$ and
$\Pi_{\mu\nu} (p)$ are given by
\begin{eqnarray}  
\Pi^{(b)} &=& {1 \over 2} 
 \partial_\beta ^2 \Pi_{\alpha\alpha}
+\partial_{\alpha,\beta} ^2 \Pi_{\alpha \beta} ^{lattice}(0), 
\label{pb}
\\
\Pi^{(c)} &=& {1 \over 2} 
 \partial_\alpha ^2 \Pi_{\alpha\alpha}^{lattice}(0) -\Pi^{(b)}.
\label{pc}
\end{eqnarray}  

The most simple term in eq.(\ref{eqn:para}) is $\Pi^{(b)}$,
which explicitly given by
\begin{eqnarray}  
\Pi^{(b)} &=& {-g^2 \over 8} 
\int_{-\pi} ^{\pi} {d^n q \over (2\pi)^n} {\rm tr} 
\left\{ 
\left[\partial_\beta ^2 L_\alpha (q) \star L_\alpha (q)\right]
\right. \nonumber \\ 
&-&  
2 \left[\partial_\beta L_\alpha (q) \star
         \partial_\beta L_\alpha (q) \right]
+\left[L_\alpha (q) \star \partial_\beta ^2 L_\alpha (q)\right]
\nonumber \\ 
&+& 
2\left[\partial_\beta L_{\alpha\alpha} (q) \star L_\beta (q)\right]
-2 \left[L_{\alpha \alpha}(q) \star L_{\beta \beta}(q) \right] \\ 
&-& \left.
2 \left[\partial_\beta L_\alpha (q) \star
         \partial_\alpha L_\beta (q) \right]
+2 \left[L_\alpha (q) \star \partial_\alpha ^2 L_{\beta\beta} (q)\right]
\right\}, \nonumber 
\end{eqnarray}  
where $[L_\mu]_{st} (q)
=[\partial_\mu S_F ^{-1}(q)\cdot S_F(q)]_{st}$ and
$[L_{\mu\mu}]_{st} (q)
=[\partial_\mu S_F ^{-1}(q)\cdot \partial_\mu S_F(q)]_{st}$.
We obtain eq.(\ref{eqn:2point-b})
by substituting the following relation into the last equation.
\begin{eqnarray}  
L_{\mu\mu}(q) =
\partial_\mu ^2 S_F ^{-1}(q)\cdot S_F(q)
+L_\mu(q)\cdot L_\mu(q). 
\label{formula1}
\end{eqnarray}  
Here we use the fact that the following n-dimensional integral is zero 
by the Stokes' theorem:
\begin{eqnarray}  
\int_{-\pi} ^\pi {d^n q \over (2\pi)^n} {\rm tr}
\partial_\beta ^2 
\left[L_\alpha (q) \star L_\alpha (q)\right] = 0.
\end{eqnarray}  
$\Pi^{(c)}$ is also obtained in the same way.
Note that both $\Pi^{(b)}$ and $\Pi^{(c)}$ have no infra-red singularity.

Next we consider the most interesting term $\Pi^{(a)}$. 
For convenience 
eq.(\ref{pa}) is rewritten with eq.(\ref{pb}) as
\begin{eqnarray}  
\Pi^{(a)} &=& -\Pi^{(b)} +{1 \over 2} 
 \partial_\beta ^2 \Pi_{\alpha\alpha}(0).
\end{eqnarray}  
By this identity we can concentrate on the last term 
for the calculation of $\Pi^{(a)}$.
The last term becomes
\widetext
\begin{eqnarray}  
&& {-g^2 \over 8} 
\int_{-\pi} ^{\pi} {d^n q \over (2\pi)^n} {\rm tr} 
\left\{ 
\left[\partial_\beta ^2 L_\alpha (q) \star L_\alpha (q)\right]
-2 \left[\partial_\beta L_\alpha (q) \star
         \partial_\beta L_\alpha (q) \right]
+\left[L_\alpha (q) \star \partial_\beta ^2 L_\alpha (q)\right]
\right\} \nonumber \\
&=& {-g^2 \over 8} 
\int_{-\pi} ^{\pi} {d^n q \over (2\pi)^n} {\rm tr} 
\partial_\beta \left\{ 
 \left[\partial_\beta L_\alpha (q) \star L_\alpha (q)\right]
-\left[L_\alpha (q) \star \partial_\beta L_\alpha (q)\right]
\right\} 
\\
&&  \qquad \qquad \qquad \qquad \qquad \qquad 
-4 \times {\rm tr} \left\{ 
 \left[\partial_\beta L_\alpha (q) \star
         \partial_\beta L_\alpha (q) \right]
\right\}.\nonumber 
\end{eqnarray}  
To derive the last equation we use the formula (\ref{formula1}) again.
By the Stokes' theorem in n dimensions,
this integral simply becomes
\begin{eqnarray}  
\Pi^{(a)}+\Pi^{(b)} &=& {g^2 \over 2}
\int_{-\pi} ^{\pi} {d^n q \over (2\pi)^n}  {\rm tr} 
\left\{ 
\left[\partial_\beta L_\alpha (q) \star
         \partial_\beta L_\alpha (q) \right]
\right\}.
\end{eqnarray}  
Since this integral has a logarithmic infrared singularity
we evaluate it in the following way:
\begin{eqnarray}  
\Pi^{(a)}+\Pi^{(b)} &=& 
\left[\Pi^{(a)} - \Pi^{(a)}|^{cont.} \right]
+ \Pi^{(a)}|^{cont.} + \Pi^{(b)} , 
\end{eqnarray}  
where $\Pi^{(a)}|^{cont.}$ means
the integrand in $\Pi^{(a)}$ is replaced by the continuum one:
\widetext
\begin{eqnarray}  
\Pi^{(a)}|^{cont.}&=& 
{-g^2 \over 8} (-4){1 \over 2}
\sum_\chi F^2_\chi (s,t) 
\int_{-\pi} ^{\pi} {d^n q \over (2\pi)^n}  
{\rm tr} \left\{
i \gamma_\alpha {-i {q\!\!\!\!/\,\,} \over q^2}
i \gamma_\beta  {-i {q\!\!\!\!/\,\,} \over q^2}
i \gamma_\alpha {-i {q\!\!\!\!/\,\,} \over q^2}
i \gamma_\beta  {-i {q\!\!\!\!/\,\,} \over q^2}
\right\} \nonumber \\
&=&
{-g^2 (-4 \cdot 4) \over 8 \cdot 2} 
  \sum_\chi F^2_\chi (s,t) 
\int_{-\pi} ^{\pi} {d^n q \over (2\pi)^n} 
\left\{ 
 {8q_\alpha ^2 q_\beta ^2 \over (q^2)^4} -{1 \over (q^2)^2}
\right\}, \nonumber 
\end{eqnarray}  
and $\Pi^{(b)}|^{cont.} = 0$.
Using a property of the Gauss's error function, we obtain
\begin{eqnarray}  
\Pi^{(a)}|^{cont.}&=& 
-g^2 \sum_\chi F^2_\chi (s,t) 
\left\{
{2 \over 3 (2\pi)^4} 
\int_0^{\infty} dy y^{1-n/2} (<1>_y)^n
+{1 \over (4\pi)^2} 
 \left( {1 \over 3} -{2 \over \pi \sqrt{3}}\right)
\right\},
\end{eqnarray}  
\narrowtext
where $<1>_y$ is defined by 
$\int_{-\pi \sqrt{y}}^{\pi \sqrt{y}} dx 1 e^{-x^2}$.
The first term in the last step has a $1/\epsilon$ divergence.
Therefore we evaluate this integral as follows:
\begin{eqnarray}  
&&\int_0^{\infty} dy y^{1-n/2} (<1>_y)^n =
 \int_0^1 dy y^{-1} (<1>_y)^4 
\nonumber \\
&+& 
\int_1^{\infty} dy y^{-1} \left\{(<1>_y)^4 -  \sqrt{\pi}^4 \right\}
+\sqrt{\pi}^4  \int_1^{\infty} dy y^{1-n/2}. \nonumber \\
&=&
{- \pi^2 \over \epsilon} + 0.46349.
\nonumber
\end{eqnarray}  
Here we set $n=4$ for the term without infra-red divergence.

Finally we reach the result eq.(\ref{eqn:2point-a}).
By considering that
the factor $a^n \bar \mu ^n$ must be multiplied 
in front of the n dimensional integral,
it can be easily understood that
$log(a^2 \bar \mu^2)$ appears in eq.(\ref{eqn:2point-a})\cite{KN}.
This is one of the most important points 
among what the final result tells us:
The $\epsilon^{-1}$ of the lattice contribution cancels 
that of the continuum one and $\bar \mu$ dependence also disappears,
so that only the ultra-violet divergence
$\log p^2a^2$ remains in the final result.
We can also see a similar kind of
the cancellation of infra-red divergences in Ref.\ \cite{KN}.

\section*{Appendix C}
\label{sec:AppC}

We investigate the anomaly in this Appendix.
Let us start with the following definition of the three point function:
\widetext
\begin{eqnarray}
\Gamma_{\mu\nu\rho}(q,p,k)(s,t,u) &=& 
{1 \over 3}
\int_{-\pi}^\pi {d^4 l\over (2\pi)^4} 
{\rm tr} 
\left\{ 
\left[\partial_\mu S^{-1}(l-q/2) S(l) \right]_{st}
\right.\nonumber \\
&&\left. 
\star 
\left[\partial_\nu S^{-1}(l+p/2) S(l+p) \right]_{tu}
 \star 
\left[\partial_\rho S^{-1}(l+(p-q)/2) S(l-q) \right]_{us}
\right\} \nonumber \\
&& +(l \rightarrow l-p)+(l \rightarrow l-q).
\nonumber
\end{eqnarray}
Substituting the above into eq.(\ref{anolat1}), 
we obtain eq.(\ref{anolat}), which is given once more below:
\begin{eqnarray}
A_{q.p}(s,t,u) &=& 
{1 \over 4!} g^3 \epsilon_{\mu\nu\rho\sigma} {1 \over 3}
\int_{-\pi}^\pi {d^4 l\over (2\pi)^4} 
\label{apanolat} \\
&&\times
3{\rm tr} \left\{ 
\left[L_\mu (l) \cdot L_\sigma (l)\right]_{st}
\star L_\nu (l)_{tu} \star L_\rho (l)_{us}
\right\} 
-\partial_\sigma
{\rm tr} \left\{ 
L_\mu (l)_{st} \star L_\nu (l)_{tu} \star L_\rho (l)_{us}
\right\}. \nonumber 
\end{eqnarray}
\narrowtext

The dependence on the extra dimension s,t,u of this anomaly 
is calculated numerically.
On other hand, we can calculate it analytically 
in special cases -- sum over s,t and sum over u.
For considering these cases, let us prove two identities,
eq.(\ref{3pou}) and (\ref{3post}).
First we show that
\begin{eqnarray}
\left\{
\begin{array}{l}
\sum_{st} \\
\sum_{u} \\
\end{array}
\right\}
& & A_{q.p}(s,t,u) = 
{1 \over 4!} g^3 \epsilon_{\mu\nu\rho\sigma} 
\int_{-\pi}^\pi {d^4 l\over (2\pi)^4} 
{1 \over 3}
\label{sumstu} \\
&& \times 
\left\{
\begin{array}{l}
-4 \partial_\sigma 
{\rm tr} \left\{ \left[ 
L_\mu (l) \cdot L_\nu (l) \cdot L_\rho (l)
\right]_{uu}\right\} \\
+2 \partial_\sigma 
{\rm tr} \left\{ 
\left[L_\mu (l) \cdot  L_\nu (l)\right]_{st}
 \star
L_\rho (l)_{ts} \right\}
\end{array}
\right\}, \nonumber
\end{eqnarray}
using the following line of identities 
to the first term of eq.(\ref{apanolat}) 
(s,t,u are suppressed):
\begin{eqnarray}
\left\{
\begin{array}{l}
 \sum_{s,t}  \\
 \sum_u      \\
\end{array}
\right\}
3 {\rm tr} 
\left\{ \left[
\left(L_\mu (l) \cdot L_\sigma (l)\right) 
\star L_\nu (l) \star L_\rho (l)
\right]\right\}
\nonumber 
\end{eqnarray}
\begin{eqnarray}
&&   =
 \left\{
\begin{array}{l}
 3 {\rm tr} 
\left\{ \left[
      L_\rho (l) \cdot L_\mu (l) 
\cdot L_\sigma (l) \cdot L_\nu (l) 
\right] \right\} \\
 3 {\rm tr} 
\left\{ 
\left[L_\mu (l) \cdot L_\sigma (l)\right] 
\star \left[L_\nu (l) \cdot L_\rho (l)\right] 
\right\} \\
\end{array}
\right\}\nonumber \\
&& =
\left\{
\begin{array}{l}
-3 
{\rm tr} 
\left\{ \left[
\partial_\rho S^{-1}(l)\cdot \partial_\mu S(l) \cdot
\partial_\sigma S^{-1}(l) \cdot \partial_\nu S(l) 
\right]
\right\}
\\
+3 
{\rm tr} 
\left\{\left[
\partial_\mu S^{-1}(l)\cdot \partial_\nu S(l) \right] 
\star \left[
\partial_\sigma S^{-1}(l) \cdot \partial_\rho S(l) 
\right]\right\}
\\
\end{array}
\right\}.
\nonumber 
\end{eqnarray} 

Since integrands in the right hand side of eq.(\ref{sumstu})
become total derivatives,
the integrals can be performed 
with the Stokes' theorem. 
Using the equation
\begin{eqnarray}
[L_\mu (l)]_{st} = i \gamma_\mu  
\times {-i {l\!\!\!/\,\,}\over l^2}
\sum_\chi F_\chi (s,t) P_\chi ,
\ \ \ \ \ \ (l \simeq 0  )
\nonumber 
\end{eqnarray}
and introducing the infra-red cut-off $\epsilon$, 
the total derivative is evaluated as
\widetext
\begin{eqnarray}
 &&
{1 \over 4!} g^3 \epsilon_{\mu\nu\rho\sigma} 
\int_{-\pi}^\pi {d^4 l\over (2\pi)^4} 
\partial_\sigma
{\rm tr} \left\{
L_\mu (l) \star L_\nu (l) \star L_\rho (l)
 \right\} 
\nonumber \\
&=&
{1 \over 4!} g^3 \epsilon_{\mu\nu\rho\sigma} 
\int_{-\pi}^\pi {d^3 \vec l \over (2\pi)^4} 
2 \times \left.
\left\{ i 4! \epsilon_{\mu\alpha\nu\rho} 
\sum_\chi {\delta_\chi \over 2}  F_\chi^3
{\sum_\alpha l_\alpha \over (l^2)^2} 
\right\} 
\right|_{l_\sigma = \epsilon} ^{l_\sigma = \pi/2}
=
g^3 \sum_\chi \delta_\chi  F_\chi^3 {-i \over 16 \pi^2}.
\nonumber
\end{eqnarray}
In the last equality we used the following formula:
\begin{eqnarray}
\int_{-\pi}^\pi {d^3 \vec l \over (2\pi)^4} 
\left.{l_\alpha \over (l^2)^2} 
\right|_{l_\sigma = \epsilon} ^{l_\sigma = \pi/2} 
&=& 
-\int_{-\pi/\epsilon}^{\pi/\epsilon} {d^3 \vec l \over (2\pi)^4} 
\int_0 ^1 dx x e^{-x(1+ \vec l ^2)}
= -{1 \over 16 \pi^2}. \ \ \ (\epsilon \simeq 0).
\label{apint2}
\end{eqnarray}
Eq.(\ref{sumstu}) and (\ref{apint2}) lead to 
identities eq.(\ref{3pou}) and (\ref{3post}).

We shall turn to 
the proof of 
$\epsilon_{\mu\nu\rho\sigma}
\partial_\sigma \Gamma_{\mu\nu\rho}(0,0,0;s,t,u)|^{cont.}= 0$
in eq.(\ref{3poreg}),  
which is nontrivial in our regularization,
using the same technique above.
The calculation is almost the same as that for eq.(\ref{apanolat})
except that the integral $ \int_{-\pi} ^{\pi}$ 
$\rightarrow \int_{-\Lambda}^\Lambda $
and $L_\mu (l)$ is that of the continuum:
\begin{eqnarray}
&&
{1\over 4!} \epsilon_{\mu\nu\rho\sigma}
\left.
\partial_\sigma \Gamma_{\mu\nu\rho}(0,0,0;s,t,u) 
\right|^{cont.} \nonumber \\
=&& {1 \over 4!} g^3 \epsilon_{\mu\nu\rho\sigma} 
{1 \over 3}
\sum_\chi F_\chi ^3
\int_{-\Lambda}^\Lambda {d^4 l\over (2\pi)^4}
3{\rm tr} \left\{ 
L_\mu (l) \cdot L_\sigma (l)
\cdot L_\nu (l) \cdot L_\rho (l)\right\} 
-\partial_\sigma
{\rm tr} \left\{ 
L_\mu (l) \cdot L_\nu (l) \cdot L_\rho (l)
\right\}  \nonumber \\
=&& {1 \over 4!} g^3 \epsilon_{\mu\nu\rho\sigma} {1 \over 3}
\int_{-\Lambda}^\Lambda {d^4 l\over (2\pi)^4}
 (-4) {\rm tr} \left\{ 
L_\mu (l) \cdot L_\sigma (l)
\cdot L_\nu (l) \cdot L_\rho (l)
\right\}. 
\nonumber
\end{eqnarray}

\narrowtext
Therefore we should calculate the following integral
instead of eq. (\ref{apint2}).
\begin{eqnarray}
&& \int_{-\Lambda}^\Lambda {d^3 \vec l \over (2\pi)^4} 
\left.{l_\alpha \over (l^2)^2} 
\right|_{l_\sigma = \epsilon} ^{l_\sigma = \Lambda} \nonumber \\
&=& 
-\int_{-\Lambda}^{\Lambda} {d^3 \vec l \over (2\pi)^4} 
\int_0 ^1 dx x e^{-x(1+ \vec l ^2)} \nonumber \\
&&\qquad -\left(-
\int_{-\Lambda/\epsilon}^{\Lambda/\epsilon} 
{d^3 \vec l \over (2\pi)^4} 
\int_0 ^1 dx x e^{-x(1+ \vec l ^2)}
\right)\nonumber \\
&=& -{1 \over 16 \pi^2}-(-{1 \over 16 \pi^2}) =0 .\nonumber 
\end{eqnarray}
This means that the derivative term in the continuum,
$\epsilon_{\mu\nu\rho\sigma}
\partial_\sigma \Gamma_{\mu\nu\rho}(0,0,0;s,t,u)|^{cont.}$,
is zero in this  regularization.

\begin{figure}
\centerline{\epsfxsize=12cm \epsfbox{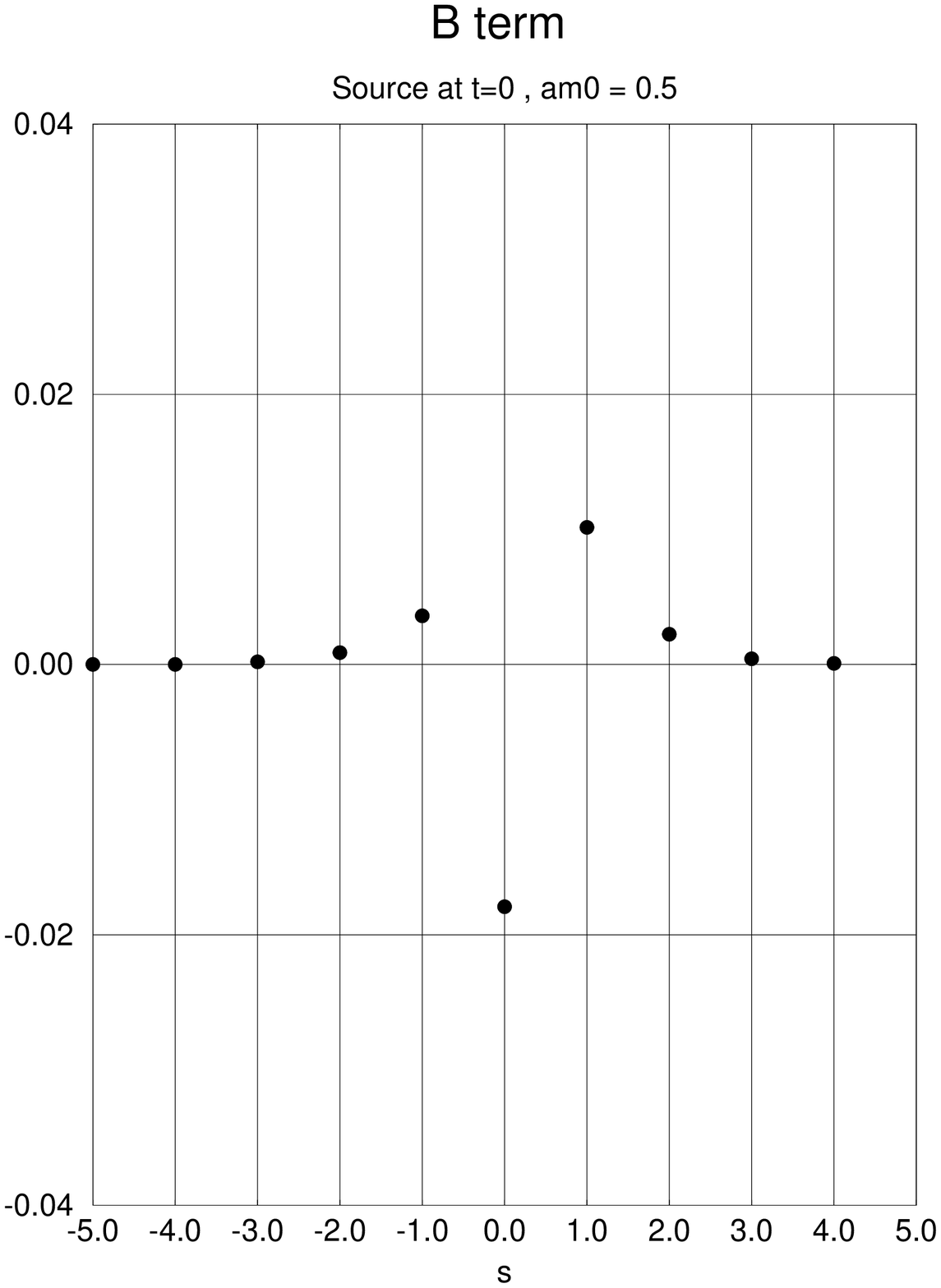}}
\caption{The coefficient of the 2 point function
$\Pi^{(b)}_{st}$ as a function of $s$ 
with $t=0$ fixed at $m_0=0.5$ and $L=5$.
}
\label{longi4}
\end{figure}

\end{document}